\documentclass[aps,pre,showpacs,groupedaddress,twocolumn]{revtex4}

\usepackage{graphicx,psfrag,epsfig}
\usepackage[hypertex]{hyperref}

\newcommand{\C}{{\mathrm c}}
\newcommand{\D}{{\mathrm d}}
\newcommand{\E}{\mathrm{e}}
\newcommand{\F}{\mathrm{f}}
\newcommand{\I}{\mathrm{i}}
\newcommand{\DD}{\mathcal{D}}
\newcommand{\FF}{\mathcal{F}}
\newcommand{\HH}{\mathcal{H}}
\newcommand{\LL}{\mathcal{L}}
\newcommand{\NN}{\mathcal{N}}
\newcommand{\Sc}{\mathcal{S}}

\newcommand{\tf}{{t_\mathrm{f}}}
\newcommand{\average}[1]{\left<{#1}\right>}

\newcommand{\pu}{p^\uparrow}
\newcommand{\pd}{p^\downarrow}

\newcommand{\p}[1]{\left({#1}\right)}
\newcommand{\pq}[1]{\left[{#1}\right]}
\newcommand{\pg}[1]{\left\{{#1}\right\}}
\newcommand{\derpart}[2]{\frac{\partial #1}{\partial #2}}

\begin{document}

\title{Work probability distribution in systems driven out of equilibrium}

\author{A. Imparato$^\dagger$}
\email[Corresponding author. Email: ]{imparato@na.infn.it}
\author{L. Peliti}\thanks{Associati INFN, Sezione di Napoli.}

\affiliation{Dipartimento di Scienze Fisiche and Unit\`a INFM,\\
Universit\`a ``Federico II'', Complesso Monte S. Angelo,
I--80126 Napoli (Italy)}

\date{July 26, 2005}

\begin{abstract}
We derive the differential equation describing the
time evolution of the work probability distribution
function of a stochastic system which is driven
out of equilibrium by the manipulation of a parameter.
We consider both systems described by their microscopic state
or by a collective variable which identifies a quasiequilibrium state.
We show that the work probability distribution can
be represented by a path integral, which is dominated by
``classical'' paths in the large system size limit.
We compare these results with simulated manipulation
of mean-field systems.
We discuss  the range of applicability of the Jarzynski
equality for evaluating the system free energy using
these out-of-equilibrium manipulations.
Large fluctuations in the work and the shape of
the work distribution tails are also discussed.
\end{abstract}


\maketitle
Recent improvements in micromanipulation techniques have made
it possible to observe experimentally  work fluctuations and to
measure the  probability distribution of the work exerted on a system
subject to external manipulation. In particular, the probability
distribution of the work has been measured in RNA pulling
experiments \cite{exp,felix0} and for micrometer-sized colloidal particles
dragged through a fluid \cite{exp1}. Usually, because of technical
limitations, this class of experiments is characterized by
time scales much faster than the typical system relaxation time.
This hinders the possibility to perform the
experiments in quasistatic conditions and thus to obtain direct
measurements of the system thermodynamic state variables.
The importance of the knowledge of work distributions in such
experiments resides in the fact that one can evaluate
the free energy difference between the final and the initial
state of the system by exploiting the Jarzynski equality
(JE)~\cite{JE,Crooks,Jarzlect}
\begin{equation}
\label{JE:eq}
    \average{\E^{-\beta W}}=\E^{-\beta \Delta F}.
\end{equation}
According to previous works~\cite{felix0}, a precise knowledge of
the tails in the distributions provides information
on how many experiments are needed in order to evaluate correctly
the free energy difference of a system using non-equilibrium
experiments. Thus, {\it a priori} estimates of $P(W)$ are
in principle needed, to evaluate the actual usefulness of this
approach.

In two recent works \cite{noi1,noi2}, we
introduced and discussed a differential equation describing the
time evolution of the probability distribution of the work done on
a system by manipulating an external field (force) $\mu$,
according to a given protocol $\mu(t)$. In particular, in ref.~\cite{noi1},
we considered the case of a system characterized by a
discrete phase space, while in ref.~\cite{noi2} we considered a
mean field system characterized by a generic equilibrium free
energy $\FF_\mu(M)$.

The aim of this paper is to extend those works,
by exploiting an approach due to Felix Ritort~\cite{felix}.
In particular, we first derive explicitly the differential
equations governing the time evolution of $P(W,t)$.  We then derive
an expression of the work probability distribution of a system
described by a collective variable, on the hypothesis that, during the
manipulation, the system finds itself in a quasiequilibrium state
constrained by the value of that coordinate. We solve the resulting
equation by path integrals and show that, in the limit of large system size,
the path integral is dominated by the classical path which satisfy
canonical equations of motion, and suitable boundary conditions.
The expression for the probability distribution function follows
straightforwardly. We highlight the analogy between the path functionals
obtained in this way and classical thermodynamics. We apply the
obtained results to some simple systems, and we explore in particular
the possibility of the existence of exponential
tails in the work probability distribution: such tails are
related, via the thermodynamic analogy, to phase transitions
in the path distribution. We show that, contrary to what was conjectured
in ref.~\cite{felix} on the basis of numerical evidence, such tails
are not present in a paramagnet, or in a ferromagnet above
the critical temperature, but are present in a mean-field ferromagnet
below the critical temperature, provided the manipulating protocol is
fast enough. The implications of our results are further discussed.

\section{Probability distribution of the work
for the microscopic coordinates}\label{sec1}
 In this section, we see how the
probability distribution function of the work $W$ exerted on a
system can be evaluated by considering the joint probability
distribution of $W$ and the microscopic state of the system.
This equation was derived in refs.~\cite{noi1,Seifert} (see also
 \cite{Jarzlect}).
Let us first consider a system whose microscopic state $i$ can take on
a finite number of values. To each such state is assigned an
energy value $H_i(\mu)$, where $\mu$ is a parameter which is
manipulated according to some protocol $\mu(t)$, starting at
$t=0$. We assume that the evolution of the system is described by
a markovian stochastic process: given, for all pairs $(i,j)$, the
transition rate $k_{ij}(t)$ from state $j$ to state $i$ at time
$t$, the system satisfies the set of differential equations
\begin{equation}
    \derpart{p_i}{t}= \sum_{j(\ne i)} \left[k_{ij}(t) p_j(t)
    - k_{ji}(t) p_i(t)\right], \label{evp}
\end{equation}
where $ p_i(t)$ is the probability that the system is found at
state $i$ at time $t$. Let $p^\mathrm{eq}_i(\mu)$ represent the
equilibrium distribution corresponding to a given value of $\mu$.
We have
\begin{equation}
    p^\mathrm{eq}_i(\mu)=\frac{\E^{-\beta H_i(\mu)}}{Z_\mu},
\end{equation}
where $Z_\mu=\sum_i \E^{-\beta H_i(\mu)}=\E^{-\beta F_\mu}$ is the partition
function corresponding to the value $\mu$ of the parameter, and $F_\mu$
the corresponding free energy. We
require that the transition rates $k_{ij}(t)$ are compatible with
the equilibrium distribution $p^\mathrm{eq}_i(\mu)$, i.e.,
that, for any $i$,
\begin{equation}
    \sum_{j (\neq i)}\left[k_{ij} (t) p^{\mathrm{eq}}_j(\mu(t))
    - k_{ji} (t) p^{\mathrm{eq}}_i(\mu(t))\right]
   =0.
\label{equil}
\end{equation}
We assume that the system is at equilibrium at $t=0$, and
therefore, that $p_i(t)$ satisfies the initial condition
\begin{equation}
    p_i(t{=}0)=p^\mathrm{eq}_i(\mu(0)).
\end{equation}
As pointed out in ref.~\cite{noi1}, the function $p_i(t)$ does
not provide sufficient information on the work performed on the
system during the manipulation process. We can however consider
the joint probability distribution $\Phi_i(W,t)$ that the system
is found in state $i$, having received a work $W$, at time $t$. If
the system is in the state $i$ at time $t$, the infinitesimal work
$\delta W_i$ done on it in the interval $\delta t$ reads
\begin{equation}
    \delta W_i=\dot \mu\, \frac{\partial H_i(\mu(t))}{\partial \mu} \,\delta t.
\label{dW}
\end{equation}
We have thus
\begin{widetext}
\begin{eqnarray}
    \Phi_i(W,t+\delta t) &\simeq &\Phi_i(W-\delta
    W_i,t)+ \delta t  \sum_{j(\ne i)} \left[k_{ij} (t) \Phi_j(W-\delta
    W_j,t)- k_{ji} (t) \Phi_i(W-\delta W_i,t) \right] \nonumber \\
   &=& \Phi_i(W,t) -\delta t\, \dot \mu \,H'_i(\mu(t))\,
    \partial_W   \Phi_i(W,t)+\delta t  \sum_{j(\ne i)}\left[ k_{ij} (t)
    \Phi_j(W,t) - k_{ji} (t) \Phi_i(W,t)\right].
    \label{dphi}
\end{eqnarray}
\end{widetext}
The last equality is obtained by substituting the expression for
$\delta W_i$ given in eq.~(\ref{dW}) and by taking the first order
expansion in $\delta t$ of the rhs. We are now able to write the
set of differential equations which describe the distribution
functions $\Phi_i(W,t)$
\begin{eqnarray}
    \derpart{\Phi_i}{t}&=&\sum_{j(\ne i)}\left[ k_{ij} (t)
    \Phi_j(W,t) - k_{ji} (t) \Phi_i(W,t)\right]\nonumber\\
    &&\qquad{}-\dot \mu H'_i(\mu(t))\,\derpart{  \Phi_i}{W}. \label{eqphi}
\end{eqnarray}
The joint probability distribution $\Phi_i(W,t)$ satisfies the
initial condition
\begin{equation}
    \Phi_i(W,0)=\delta(W)\,p^\mathrm{eq}_i(\mu(0)).
    \label{ivc}
\end{equation}

We are interested in the state-independent
work probability distribution $P(W,t)$ defined by
\begin{equation}
    P(W,t)=\sum_i\Phi_i(W,t).
    \label{eqpw}
\end{equation}
It is convenient to introduce the generating function of $\Phi_i$
with respect to the work distribution, defined by
\begin{equation}
    \Psi_i(\lambda,t)=\int \D W \;\E^{\lambda W} \Phi_i(W,t).
\label{defpsi}
\end{equation}
(Notice that we adopt here, for later convenience,
the opposite sign convention
with respect to that adopted in ref.~\cite{noi2}.)
We assume that $\Phi_i(W,t)$ vanishes fast enough, as
$|W|\to\infty$, for $\Psi_i(\lambda,t)$ to exist for any
$\lambda$. The function $\Psi_i$ satisfies the initial  condition
\begin{equation}
    \Psi_i(\lambda,t_0)=\frac{\exp\pq{-\beta
    H_i(\mu(0))}}{Z_{\mu(0)}},
\label{incpsi}
\end{equation}
and evolves according to the differential equation
\begin{eqnarray}
   && \partial_t \Psi_i(\lambda,t) =
   \int \D W\; \E^{\lambda W} \partial_t\Phi_i(W,t)\nonumber \\
   && {}= \int \D W\; \E^{-\lambda W}
   \left\{ \sum_{j(\ne i)} \pq{k_{ij} \Phi_j -k_{ji} \Phi_i}
   -\dot \mu \derpart{}{\mu}\derpart{  \Phi_i}{W}\right\}\nonumber \\
   && {}=\sum_{j(\ne i)} \pq{k_{ij} \Psi_j -k_{ji} \Psi_i}+\lambda \dot
   \mu\,\frac{\partial H_i(\mu(t))}{\partial \mu}\,
   \Psi_i(\lambda, t). \label{detpsi}
\end{eqnarray}
Exploiting eq.~(\ref{equil}), it is easy
to verify that if $\lambda=-\beta$, for any $i$ at any time $t$,
the solution of eq.~(\ref{detpsi}), with the initial condition
(\ref{incpsi}), reads
\begin{equation}\label{psimt}
    \Psi_i(-\beta, t)=\frac{\E^{-\beta H_i(\mu(t))}}{Z_{\mu(0)}}
    =\frac{Z_{\mu(t)}}{Z_{\mu(0)}}\,p^\mathrm{eq}_i(\mu(t)).
\end{equation}
We can thus straightforwardly verify the Jarzynski equality:
\begin{eqnarray}
    \average{\E^{-\beta W}}&=& \int \D W\;\E^{-\beta W}
    P(W,t)\nonumber\\
    &=&\sum_i\int \D W\;\E^{-\beta W}\,\Phi_i(W,t)\nonumber\\
    &=&\sum_i
    \Psi_i(-\beta,t)=\frac{Z_{\mu(t)}}{Z_{\mu(0)}}
    \sum_i p^\mathrm{eq}_i(\mu(t))\nonumber\\
    &=&\frac{Z_{\mu(t)}}{Z_{\mu(0)}}
    =\E^{-\beta\left(F(\mu(t))-F(\mu(0))\right)}.\label{jarzder}
\end{eqnarray}
It is thus possible, in principle, to evaluate the probability
distribution function of the work $W$ by solving the equations
(\ref{eqphi}) or (\ref{detpsi}) for all the microscopic states
$i$. This approach has been implemented in ref.~\cite{noi1} for a
simple model of a biopolymer.

\section{Collective variables}\label{sec2}
The approach discussed in the previous section becomes quickly
unwieldy as the complexity of the system increases: the dimension
of the system (\ref{eqphi}) is equal to the number of microscopic
states of the system. Clearly the system phase space must be
sufficiently small for this protocol to be carried out, as in the
case discussed in \cite{noi1}. In all the other cases, where the
system considered is characterized by a large number of degrees of
freedom, one usually introduces some collective variables, and
an effective free energy, in order to reduce the complexity of the
problem. The assumption underlying this approach is that the
system reaches on a comparatively short time scale a
quasiequilibrium state constrained by the instantaneous value of
the collective coordinate. Thus, on the the time scale of the
experiment, the state of the system can be well summarized by the
collective coordinate, with the corresponding free energy playing
the role of the hamiltonian.

Thus, we consider in the following a system
characterized by a generic equilibrium free energy function
$\FF_\mu(M)$, where $\mu$ is again the parameter which is
manipulated, and $M$ is some collective (mean-field) variable. (We
shall consider in the following the case in which $M$ is a scalar,
but the analysis holds also if $M$ is a collection of real
variables.) We assume that the system dynamics is stochastic and
markovian: let $P(M,t)$ denote the probability distribution
function of the variable $M$ at time $t$, then its time evolution
will be described by the differential equation
\begin{equation}
    \derpart{P}{t}=\widehat\HH\,P,
\end{equation}
where $\widehat\HH$ is a differential operator which depends
on the parameter $\mu$. We require that the operator
$\widehat\HH$ is compatible with the equilibrium distribution
function of the system, i.e., that the relation
\begin{equation}\label{equilibrium}
    \widehat\HH\, \E^{-\beta \FF_\mu(M)}=0
\end{equation}
holds for any value of $\mu$.

The developments which follow were first obtained
in ref.~\cite{felix} for a collection of noninteracting spins.

We will consider a general mean-field system, described by a collective variable
$M$ and a generic free energy function $\FF_\mu(M)$. (The derivation
can be easily generalized to the case in which $M$ has more
than one component.)
The work done on a system during the manipulation, along a given
stochastic trajectory $M(t)$, is given by
\begin{equation}
    W=\int_0^{t} \D t'\,  \dot \mu(t') \,
    \frac{\partial \FF_\mu(M(t'))}{\partial \mu}\, .
\end{equation}
Using the same arguments as for the discrete case, one finds that
the time evolution of the joint probability distribution
$\Phi(M,W,t)$ of $M$ and $W$ is described by the differential
equation
\begin{equation}
    \label{phi:eq}
    \frac{\partial \Phi}{\partial t}=\widehat \HH \Phi
    -\dot \mu \frac{\partial \FF_\mu}{\partial \mu}
    \frac{\partial\Phi}{\partial W},
\end{equation}
It can be easily shown that the solution of eq.~(\ref{phi:eq})
satisfies the Jarzynski equality (\ref{JE:eq})
identically~\cite{noi2}.

Equation (\ref{phi:eq}) becomes much easier to treat if one
introduces the generating function $\Psi(M,\lambda,t)$ for the
work distribution:
\begin{equation}
    \label{psi:def}
    \Psi(M,\lambda,t)=\int \D W\,\E^{\lambda W}\,\Phi(M,W,t).
\end{equation}
Equation (\ref{phi:eq}) becomes thus
\begin{equation}
    \label{psi:eq}
    \frac{\partial \Psi}{\partial t}=\widehat \HH \Psi
    +\lambda\dot \mu \frac{\partial \FF_\mu}{\partial \mu}\Psi,
\end{equation}
with the initial condition
\begin{equation}
    \Psi(M,\lambda, 0)=\frac{\E^{-\beta \FF_{\mu(0)}(M)}}{Z_{\mu(0)}}.
\label{incond}
\end{equation}
These equations are exact for a collection of free spins,
or for a mean-field Ising model. The partial
differential equation (\ref{psi:eq}) replaces the $2^N$ ordinary
differential equations (\ref{detpsi}), with $i\in\{-1,+1\}^N$,
that one would obtain without the use of the collective coordinate $M$.

We now derive a path integral representation of the solution of
eq.~(\ref{psi:eq}), taking for the differential operator
$\widehat\HH$ the expression
\begin{equation}
\label{diffop:def}
    \widehat \HH\cdot{}=\sum_{k=0}^\infty \frac{\partial^k}{\partial M^k}
    \left\{g_k(M)\cdot{}\right\}.
\end{equation}
(The coefficients $g_k(M)$ also depend on $\mu$, but this
dependence is understood to lighten the notation.) Let us
introduce the generating function of $\Psi(M,\lambda,t)$:
\begin{equation}\label{omega:def}
    \Omega(\gamma,\lambda,t)=\int \D M \;\E^{-\gamma M}\Psi(M,\lambda,t) .
\end{equation}
Multiplying both sides of eq.~(\ref{psi:eq}) by $\exp(-\gamma M)$,
and integrating over $M$, we obtain
\begin{eqnarray}
    &&\partial_t \Omega(\gamma,\lambda,t)=\int \D M \;\E^{-\gamma M}
    \p{\widehat \HH +\lambda\dot \mu \,\partial_\mu \FF_\mu} \Psi\nonumber\\
    &&\qquad{}=\int \D M \;\E^{-\gamma M}
    \left[ \sum_k \frac{\partial^k}{\partial M^k}
    \p{g_k  \Psi}+\lambda\dot \mu\,
    \partial_\mu \FF_\mu  \Psi\right]\nonumber\\
    &&\qquad{}=\int \D M \;\E^{-\gamma M}
    \left[ \sum_k\gamma^k g_k +\lambda\dot \mu\,
      \partial_\mu \FF_\mu  \right]\Psi.
\end{eqnarray}
Then the function $\Omega(\gamma, \lambda,t)$ satisfies
\begin{eqnarray}
    &&\Omega(\gamma, \lambda,t+\delta t)=\int \D M\; \E^{-\gamma M}\nonumber \\
    &&\qquad \pg{1+\delta t\pq{\HH(\gamma,M)+\lambda\dot \mu\,
      \partial_\mu \FF_\mu}  } \Psi,
\end{eqnarray}
where
the function $\HH(\gamma,M)$ is defined as
\begin{equation}
    \HH(\gamma,M)=\sum_k \gamma^k g_k(M).
    \label{hgm}
\end{equation}
Given $\Omega(\gamma, \lambda,t)$, we can evaluate
$\Psi(M,\lambda,t)$ from the expression
\begin{equation}
    \Psi(M,\lambda,t)=\int^{+\I \infty}_{-\I \infty} \frac{\D
    \gamma}{2\pi \I} \;\E^{\gamma M}\, \Omega(\gamma, \lambda,t).
\end{equation}
(In the following, we shall understand the integration limits on
$\gamma$.) We obtain therefore
\begin{eqnarray}
    &&\Psi(M,\lambda,t+\delta t)
    =\int \frac{\D \gamma}{2\pi \I} \int \D M'\;\E^{\gamma (M-M')}\nonumber \\
    & &\qquad\pg{1+\delta t\pq{\HH(\gamma,M')
    +\lambda\dot \mu\,  \partial_\mu \FF_\mu}}  \Psi(M',\lambda,t)\nonumber\\
    &&\quad {}\simeq\int \frac{\D \gamma}{2\pi \I}
    \int \D M'\;\E^{\gamma (M-M')
    +\delta t \pq{\HH(\gamma,M')+\lambda\dot \mu\,
      \partial_\mu \FF_\mu}}\nonumber \\
    &&\qquad \qquad {}\times  \Psi(M',\lambda,t).
\end{eqnarray}
Iterating, we obtain
\begin{eqnarray}
    &&\Psi(M,\lambda,t+N_t \delta t)= \int \D M_0 \int
    \prod_{i=0}^{N_t}\frac{\D \gamma_i \D M_i}{2 \pi \I}\; \delta
    (M-M_t)\nonumber\\
    &&\qquad {}\times \exp\left\{\Sc[\gamma,M]\right\}\,\Psi(M_0,\lambda,0),
    \label{pathint}
\end{eqnarray}
where the ``action'' $\Sc[\gamma,M]$ is given by
\begin{eqnarray}
    &&\Sc[\gamma,M] =\sum_{i=1}^{N_t} \left\{\gamma_i (M_i
    -M_{i-1})\right.\\
    &&\qquad\qquad\qquad\left.{}+\delta t \pq{\HH(\gamma_i,M_i)
      +\lambda\dot\mu\,
    \partial_\mu \FF_{\mu(t_i)}(M_i)}\right\}. \nonumber
\end{eqnarray}
In the continuum limit, eq.~(\ref{pathint}) becomes
\begin{eqnarray}
\label{integral:def}
    \Psi(M,\lambda,\tf)&=&\int\D M_0
    \int_{M(0)=M_0}^{M(\tf)=M} \DD\gamma\DD M\;\nonumber\\
    &&\exp\left\{\Sc[\gamma,M]\right\}\;\Psi(M_0,\lambda,0),
\label{path}
\end{eqnarray}
where
\begin{equation}
    \Sc[\gamma,M]=\int_0^\tf\D t\;\LL(t).
\end{equation}
The ``lagrangian'' $\LL$ is given by
\begin{equation}
    \LL(t)=\left.\left(\gamma \dot M+\HH(\gamma,M)
    +\lambda \dot \mu \,
    \frac{\partial \FF_\mu}{\partial \mu}\right)
    \right|_{\gamma(t),M(t),\mu(t)}.
\label{LL}
\end{equation}
Let $N$ indicate the size of the system, and let us define the
``intensive quantity'' $m=M/N$. We can thus define, in the
thermodynamic limit $N\to\infty$, $m=\mbox{const.}$, the densities
\begin{eqnarray}
    f_\mu(m)&=&\lim_{N\to\infty}\frac{\FF_\mu(N m)}{N},\\
    H(\gamma,m)&=&\lim_{N\to\infty}\frac{\HH(\gamma, Nm)}{N}.\label{smallH}
\end{eqnarray}
The Lagrangian density ``per spin'' then reads
\begin{equation}
    \ell(t)=\lim_{N\to\infty}\frac{\LL(t)}{N}=\gamma \dot m+H(\gamma,m)
    +\lambda \dot \mu \,
    \frac{\partial f_\mu}{\partial \mu}.
    \label{smalll}
\end{equation}
In this way, the path integral appearing in eq.~(\ref{path})
assumes a form suitable for a saddle-point approximation for
large system sizes $N$, as pointed out in \cite{noi2,felix}.
The parameter $N$ plays a role akin to the inverse of Planck's
constant $\hbar$ in the quasiclassical approximation of
Feynman's path integral for quantum amplitudes~\cite{FeynHib}.
The result is the leading term in an asymptotic
expansion in powers of $N^{-1}$, which corresponds to the mean-field
solution of a statistical model. In ref.~\cite{felix} it was shown
that the approximation works well for free spins. In ref.~\cite{noi2}
it was shown that for a mean-field spin system above the
phase transition the approximation works rather well for
system sizes $N$ of the order of 10 and larger, but deteriorates
as the transition is approached. It would be interesting to
investigate in full the behavior of a finite-size system, in
a situation when the corresponding infinite-sized system exhibits
a phase transition. For a sufficiently fast manipulation protocol,
in a large but finite system, the probability that a
fluctuation overcoming the free energy barrier spontaneously arises
should be very small. We expect therefore that the results
of the infinite-size limit should hold better for faster
protocols than for slower ones. These issues will be dealt with in
future work.

In the leading approximation,
the path integral in eq.~(\ref{path}) is dominated by the
classical path $(\gamma_\C(t),m_\C(t))$, solution of the
differential equations
\begin{eqnarray}
    \frac{\delta \Sc}{\delta \gamma(t)}=0&\Longrightarrow& \dot m
    =-\frac{\partial H}{\partial \gamma};\label{eq1}\\
    \frac{\delta \Sc}{\delta m(t)}=0&\Longrightarrow& \dot
    \gamma=\frac{\partial H}{\partial m}+\lambda \dot \mu
    \frac{\partial^2 f_\mu}{\partial m\partial \mu}.\label{eq2}
\end{eqnarray}
We shall now see that the requirement that the system is in
equilibrium before the manipulation starts, imposes an initial
condition on these equations. In order to evaluate the integral
over $M_0$ in eq.~(\ref{path}) with the saddle-point method, we
note that $\Psi(M,\lambda,0)$ appearing on its rhs, is given by
eq.~(\ref{incond}). Furthermore, from the definition of $\ell(t)$,
eq.~(\ref{smalll}), it follows that
\begin{equation}
    \int_0^\tf \D t\,  \ell(t)=m_\tf \gamma_\tf-m_0 \gamma_0
    +\int_0^\tf \D t\, \pq{-\dot \gamma m+ H +\lambda \dot \mu
\partial_\mu f_\mu}.
\label{cambv}
\end{equation}
Thus, substituting eq.~(\ref{cambv}) into (\ref{path}), and
taking the derivative with respect to $m_0=M_0/N$,  we obtain the
saddle-point condition
\begin{equation}
\gamma(t{=}0)=-\beta \left.\frac{\partial f_\mu}{\partial m}\right|_{t=0}.
\label{gamma0}
\end{equation}

In this way one can devise a strategy to evaluate
$\Psi(M,\lambda,\tf)$ for a given manipulation protocol $\mu(t)$,
when the system size $N$ is large enough. One has to solve the
classical evolution equations (\ref{eq1},\ref{eq2}) with a
two-point boundary condition: namely, eq.~(\ref{gamma0}) should be
imposed at $t=0$, and the condition $Nm(\tf)=M$ should be imposed at
the final time $\tf$. Once the relevant classical path
$(\gamma_\C(t),m_\C(t))$ has been evaluated, one can obtain
the action density
$s[\gamma_\C,m_\C]=\lim_{N\to\infty}\Sc[\gamma_\C,N
m_\C]/N$ from the expression
\begin{equation}
    s[\gamma_\C,m_\C]=\int_0^\tf\D t\;\ell(t).
\end{equation}
Then, taking into account the initial condition (\ref{incond}),
we obtain the following asymptotic expression for
$\Psi(Nm,\lambda,\tf)$:
\begin{equation}
    \Psi(Nm,\lambda,\tf) \propto
    \exp\left\{N\left[s[\gamma_\C,m_\C]-\beta
    f_{\mu(0)}(m_\C(t{=}0))\right]\right\}.
\end{equation}
However, we are essentially interested in the state-independent
work probability distribution
\begin{equation}
    P(W,\tf)=\int \D \lambda\,  \E^{-\lambda W} \,\Gamma(\lambda,\tf),
    \label{pw}
\end{equation}
where we have defined
\begin{equation}
    \Gamma(\lambda,\tf)=\int \D M\; \Psi(M,\lambda,\tf).
\end{equation}
We shall now see that evaluating $\Gamma(\lambda,\tf)$ identifies a
well-defined boundary condition on $\gamma_\C (\tf)$. We have
indeed
\begin{eqnarray}
    &&\Gamma(\lambda,\tf) = \int \D M\, \D M_0 \int_{M(0)=M_0}^{M(\tf)=M}
  \DD\gamma\DD M\nonumber\\
    &&\qquad {}\times\exp\pq{N \int \D t\, \ell(t)}
    \Psi(M_0,\lambda,0). \label{defg}
\end{eqnarray}
In order to evaluate the integral over $M$ with the saddle point
method, we notice that, upon derivation of the rhs of
eq.~(\ref{cambv}) with respect to $m_\tf$, we obtain the condition
\begin{equation}
    \gamma_\F\equiv\gamma(\tf)=0.
\label{gammat}
\end{equation}
Thus, the equation of motions (\ref{eq1}) and (\ref{eq2})  have to
be solved with the initial and the final conditions (\ref{gamma0})
and (\ref{gammat}): let $(\gamma_\C^*(t),m_\C^*(t))$ denote
the solution of eqs.~(\ref{eq1},\ref{eq2}) satisfying these
conditions. For each value of $\lambda$, taking into account the
initial value condition (\ref{incond}), the following saddle point
estimation for $\Gamma(\lambda,\tf)$ is obtained by
eq.~(\ref{defg}):
\begin{equation}
    \Gamma(\lambda,\tf)\propto \frac{\exp\pg{-N g(\lambda)}} {Z_0},
\end{equation}
where
\begin{equation}
    g(\lambda)=\beta f_{\mu_0}(m_0^*)-\int_0^\tf \D t\;\ell^*_\C(t).
    \label{defgibbs}
\end{equation}
In this equation, $\ell^*_\C(t)$ is $\ell(t)$ evaluated along the
classical path $(\gamma^*_\C(t),m^*_\C(t))$. In order to evaluate
the integral on the rhs of eq.~(\ref{pw}), we use the saddle point
method again, and obtain
\begin{equation}
    P(Nw,\tf)=\NN\exp\pg{-N \pq{\lambda^*(w) w +g(\lambda^*(w))}},
    \label{pdg}
\end{equation}
where $\lambda^*(w)$ is the solution of
\begin{equation}\label{eqw:eq}
    g'(\lambda^*)=-w,
\end{equation}
and $\mathcal N$ is a normalization constant. Notice that the
saddle point estimate for $P(W,\tf)$ obtained in this way, implies
that the distribution becomes more and more sharply peaked around
its maximum value as $N\rightarrow\infty$. This is compatible with
the expectation that the work fluctuations becomes relatively
smaller as the size of the system increases, and in the limit
$N\to\infty$, which can be thought as the limit of a macroscopic
system, no work fluctuations are observed, and the work done on
the system during the manipulation takes one single value,
corresponding to the most probable value of $P(W,\tf)$.

In ref.~\cite{noi2} we showed that the JE is identically satisfied at the level
of classical paths. For completeness, this derivation is reproduced
in the \hyperlink{appendix}{Appendix}.

\section{A mean-field system with Langevin dynamics}\label{sec3}
We wish to discuss a few properties of the work distribution
obtained by the present method by considering a definite example.
The case of free Ising spins has been considered (within a
slightly different formalism) in ref.~\cite{felix}.
We shall return to it in sec.~\ref{expo:sec}.
We thus take an Ising-like system with mean-field interaction, with free energy
\begin{equation}
\label{freeen:eq}
    \FF(M)=-\frac{J}{2N}M^2-hM-TS(M),
\end{equation}
where where $S(M)$ is the usual entropy for an Ising paramagnet,
\begin{eqnarray}
  S(M)&=&-k_\mathrm{B}\left[\left(\frac{N+M}{2}\right)
    \log\left(\frac{N+M}{2}\right)\right.\\
    &&\qquad\qquad\left.{}+\left(\frac{N-M}{2}\right)
    \log\left(\frac{N-M}{2}\right)\right],\nonumber
  \label{entropy:eq}
\end{eqnarray}
expressed as a function of the continuous variable $M$.
We assume that the system evolves according to Langevin dynamics. The
corresponding Fokker-Planck differential operator reads
\begin{equation}
\label{FP}
    \widehat \HH \cdot {}=\omega_0N\,\frac{\partial}{\partial M}\left[
    \left(\frac{\partial \FF}{\partial M}\right)\cdot{}
    +\beta^{-1}\frac{\partial}{\partial M}\cdot{}\right],
\end{equation}
leading to the hamiltonian
\begin{equation}
  H(\gamma,m)=\omega_0\left[\gamma \left(\frac{\partial f}{\partial m}\right)
    +\beta^{-1} \gamma^2\right],
\end{equation}
where the free energy density $f(m)$ is given by
\begin{eqnarray}
  f(m)&=&-\frac{J}{2}m^2-hm+\beta^{-1}\left[\left(\frac{1+m}{2}\right)
    \log\left(\frac{1+m}{2}\right)\right.\nonumber\\
    &&\qquad\qquad\left.{}+\left(\frac{1-m}{2}\right)
    \log\left(\frac{1-m}{2}\right)\right].
\end{eqnarray}
The stochastic process described by this operator can be simulated
by integrating the corresponding Langevin equation, using the Heun
algorithm~\cite{Greiner}: for each realization of the process, the
work $W$ done on the system can be evaluated. The resulting
histogram of $w$ represents an estimate of the work probability
distribution. This estimated distribution can be then compared
with the expected distribution (valid asymptotically for
$N\to\infty$) obtained by the classical paths.
\begin{figure}[ht]
\psfrag{w}{\large $w$}
\psfrag{pw}[cb][cb][1.5]{$P(w), \, \hat P(w)$}
\psfrag{a}{$(a)$}
\psfrag{b}{$(b)$}
 \includegraphics[width=8cm]{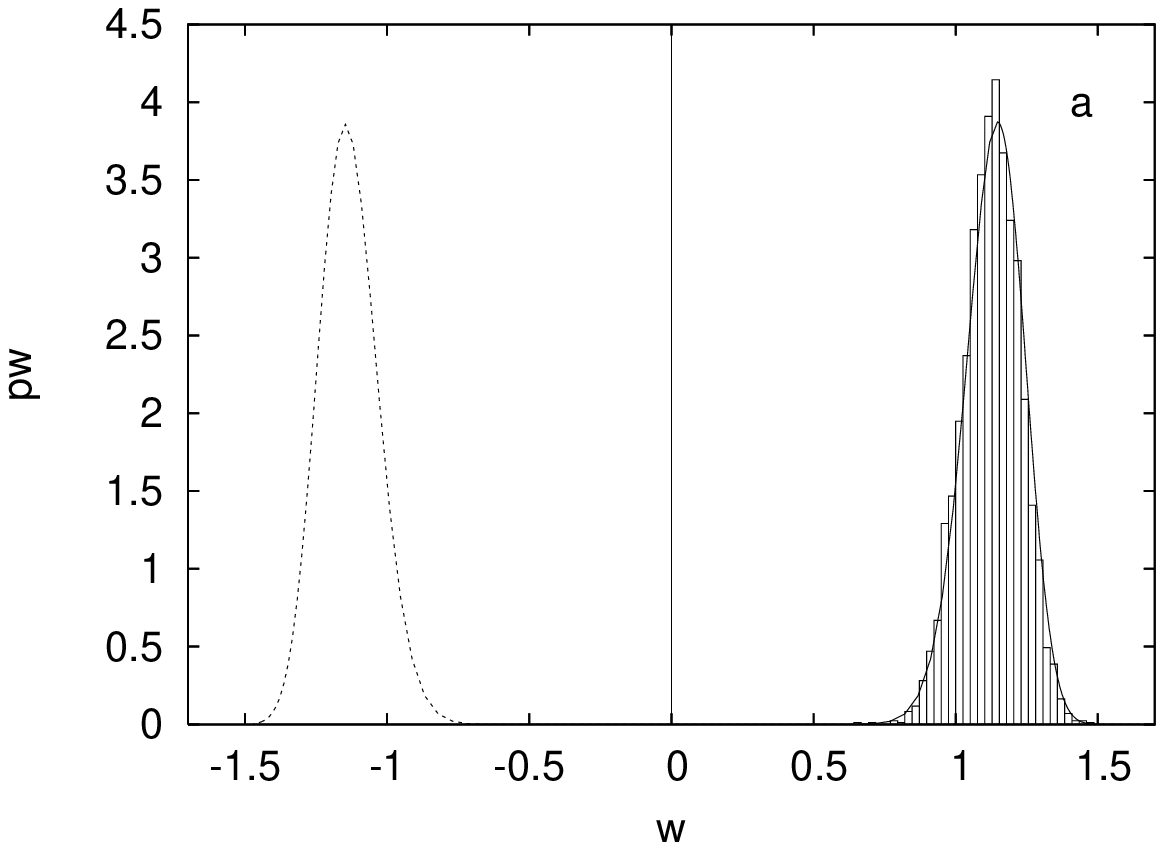}
 \includegraphics[width=8cm]{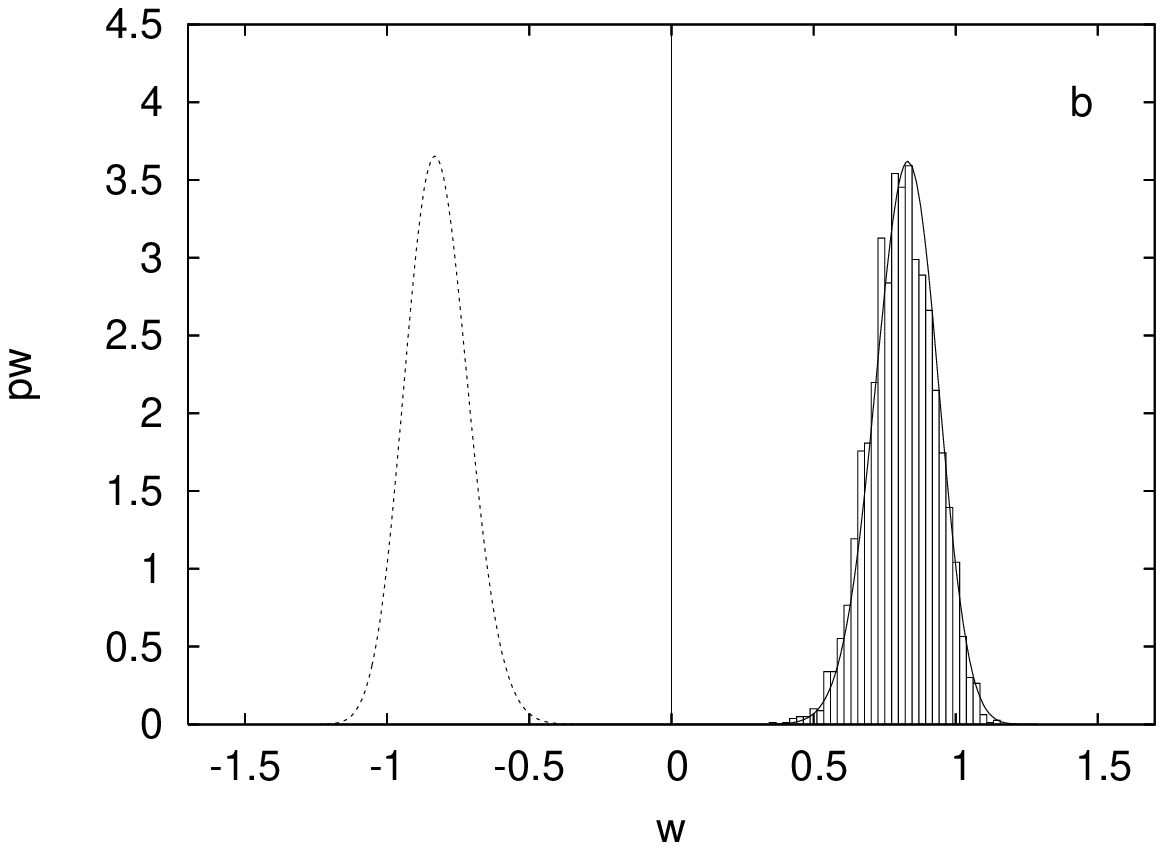}
 \caption{Results for the system described by
the differential operator (\ref{FP}) with equilibrium free energy
(\ref{freeen:eq}), manipulated according to the protocol
(\ref{manh:eq}), with $J=0.5$, $h_0=-h_1=-1$, and $(a)\,
t_\mathrm{f}=2$, $(b)\, t_\mathrm{f}=4$ .  Continuous line:
probability density $P(w)$ of the work ``per spin'' $w=W/N$, with
$N=100$. The histogram of the work is obtained by 10000
simulations of the process, see text. Dotted line:  $\hat P(w)$
as given by eq.~(\ref{hatp}), whose integral verifies
the Jarzynski equality. Vertical line: Thermodynamic value of the
work $w_\mathrm{rev}=\Delta F/N$. \label{manh}}
 \end{figure}

We consider the case where the system is subject to the
external manipulation of the magnetic field $h(t)$, according to
the simple protocol
\begin{equation}
\label{manh:eq}
h(t)=h_0+(h_1-h_0)\frac{t}{\tf}; \qquad 0\le t \le \tf.
\end{equation}
The equations of motion (\ref{eq1},\ref{eq2}) become
\begin{eqnarray}
    \dot m&=&-\derpart{H}{\gamma}=-\omega_0 \derpart{f}{m}-2 k_B T
        \omega_0\gamma,\label{eq1_lang}\\
    \dot \gamma&=&\derpart{H}{m}+\lambda \dot \mu \frac{\partial^2
    f}{\partial m\partial \mu}=\omega_0\frac{\partial^2f}{\partial
    m^2}\gamma-\lambda \dot h, \label{eq2_lang}
\end{eqnarray}
In the following we will take $\beta=1$.
In figure~\ref{manh}, we consider the case where the system is above
the critical temperature, i.e., $\beta J<1$. In this case, as
expected, the peak of the distribution moves towards the value of
the work done on the system along a reversible trajectory
$w_\mathrm{rev}=\delta f=0$, as the transformation becomes slower.
But the most important indication emerging from such a figure, is
that the JE cannot be applied to obtain an independent estimation
of the free energy difference between the final and initial states
of the transformation, if $N$ is too large. In fact, we plot in
the same figure, the quantity $\hat P(w)$ defined as
\begin{equation}
\hat P(w)=\exp\pq{-\beta N w} P(w),
\label{hatp}
\end{equation}
on the
one hand we find $\int \D w \hat P(w)=\exp\pq{-\beta \Delta F}=1$
as predicted by the JE, while on
the other hand the histogram obtained by the simulations exhibits
no point (no realization of the process) with $w<0=w_\mathrm{rev}$.
Thus the work
distribution obtained by the simulation of the process cannot
reliably be used for estimating $\Delta F$. This is a typical
example of how the lack of knowledge of the tails of the work
distributions in micro-manipulations experiments hinders the
possibility of using eq.~(\ref{JE:eq}) to evaluate free energy
differences.

\begin{figure}[hb]
\psfrag{w}{\large $w$}
\psfrag{pw}[cb][cb][1.5]{$P(w), \, \hat P(w)$}
\psfrag{a}{$(a)$}
\psfrag{b}{$(b)$}
 \includegraphics[width=8cm]{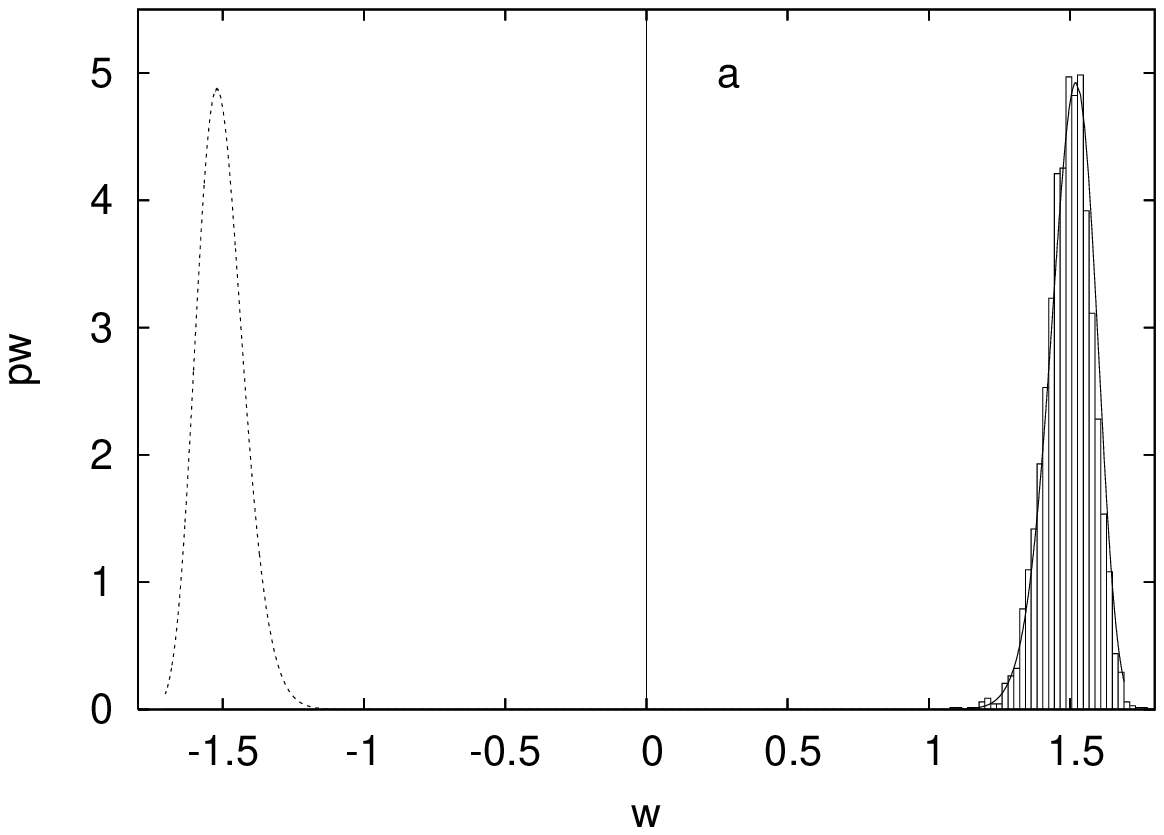}
 \includegraphics[width=8cm]{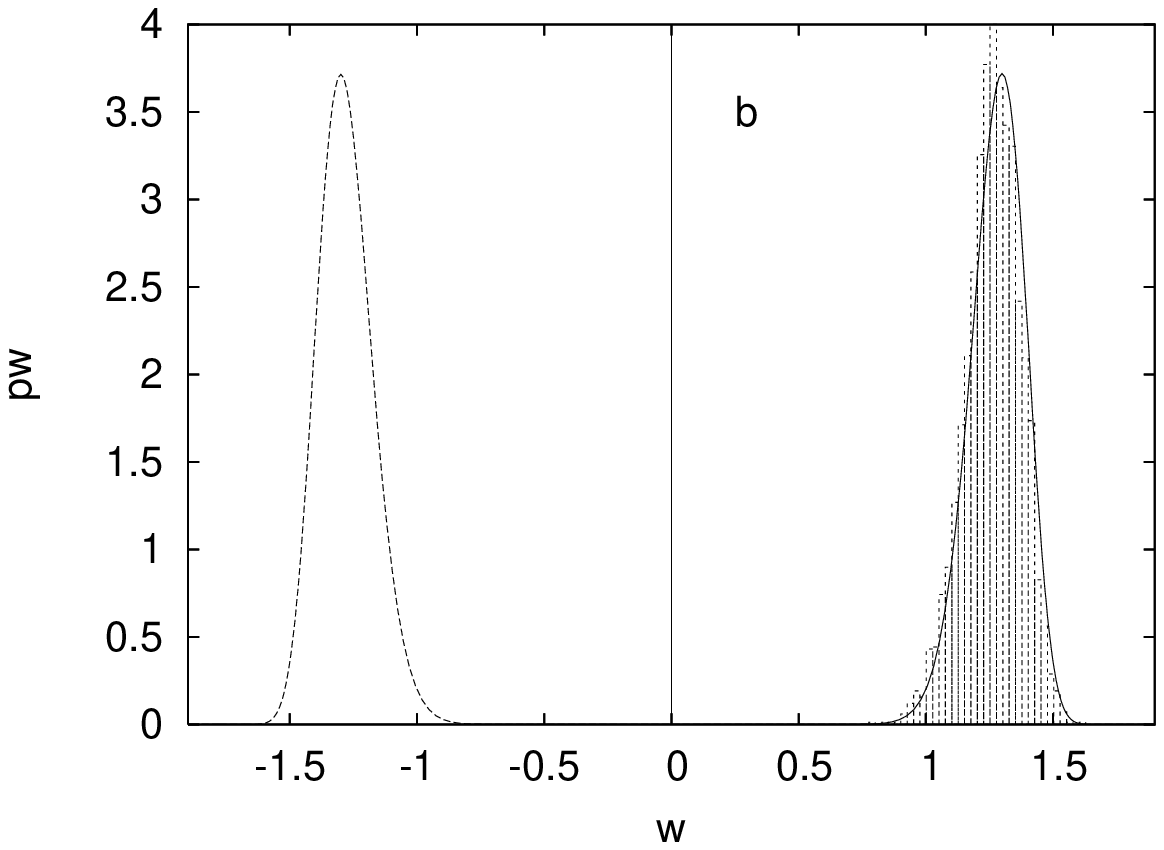}
 \caption{Results for the system described by
the differential operator (\ref{FP}) with equilibrium free energy
(\ref{freeen:eq}), manipulated according to the
protocol (\ref{manh:eq}), with $J_0=J_1=1.1$, $h_0=-h_1=-1$,
and (a)$t_\mathrm{f}=2$, (b)$t_\mathrm{f}=4$.  Continuous
line: probability density $P(w)$ of the work ``per spin''
$w=W/N$, with $N=100$.
The histogram of the work is obtained by 10000 simulations
of the process, see text. Dotted line:
 $\hat P(w)$ as given by eq.~(\ref{hatp}), whose
integral verifies the Jarzynski equality.
Vertical line: Thermodynamic value of the work
$w_\mathrm{rev}=\Delta F/N$.}
\label{manh1}
 \end{figure}

We now consider a system below the transition
temperature, i.e., for $\beta J>1$.
In figure \ref{manh1} the work probability distribution
obtained by the theory here discussed,
is plotted for $J=1.1$, $h_0=-h_1=-1$, and for two values of the final
time $\tf$.
In the same figure, the probability distribution
obtained by simulations is also plotted. As for the case
$\beta J<1$ (fig.~\ref{manh}), the JE is satisfied, i.e.,
$\average{\exp\p{-\beta W}}=1$,
there is a good agreement between the theory
and the histograms obtained by simulations.
But also in this case, such simulations cannot be used
for estimating $\Delta F$, since the histograms
exhibits no point with $w<w_\mathrm{rev}$.

\begin{figure}[ht]
\center
\psfrag{w}{\large $w$}
\psfrag{pw}[cb][cb][1.5]{$P(w), \, \hat P(w)$}
\includegraphics[width=8cm]{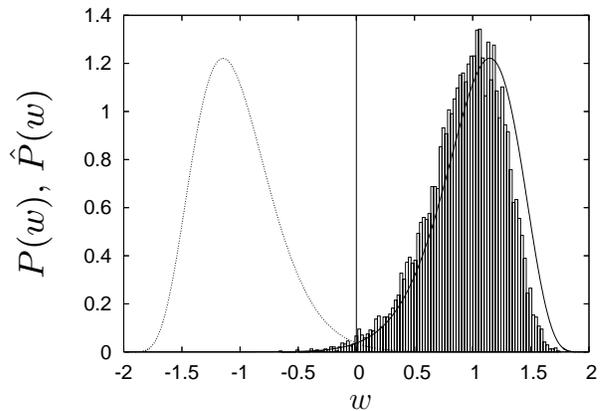}
\caption{Results for the system described by
the differential operator (\ref{FP}) with equilibrium free energy
(\ref{freeen:eq}), manipulated according to the
protocol (\ref{manh:eq}), with $J_0=J_1=0.5$, $h_0=-h_1=-1$,
and $t_\mathrm{f}=2$.  Continuous
line: probability density $P(w)$ of the work ``per spin'' $w=W/N$, with $N=10$.
The histogram of the work is obtained by 10000 simulations
of the process, see text. Dotted line: $\hat P(w)$
as given by eq.~(\ref{hatp}), whose
integral verifies the Jarzynski equality.
Vertical line: Thermodynamic value of the work
$w_\mathrm{rev}=\Delta F/N$. }
\label{n10}
\end{figure}

Since the amplitude of work fluctuations is expected
to be relatively large in small system, we calculate
now the work probability distribution for smaller
systems and compare them with the results of simulations.
First, we consider the case $N=10$, fig.\ref{n10}:
it can be seen that the
the histogram of the work obtained by simulations
is closer to the thermodynamic value of the work
$w_\mathrm{rev}=0$, than the distribution function
obtained by the theory discussed in the present paper.
Indeed, since $P(N w,t)$, as given by eq.~(\ref{pdg}),
is exact only in the limit $N\rightarrow\infty$,
that expression fails to describe the actual work
distribution for small $N$. Furthermore, even for $N=10$,
there are few points in the histogram with
$w<w_\mathrm{rev}$, and thus no reliable estimate of $\Delta F$
can be obtained from the simulations.

\begin{figure}[ht]
\center
\psfrag{w}{\large $w$}
\psfrag{pw}[cb][cb][1.5]{$P(w), \, \hat P(w)$}
\includegraphics[width=8cm]{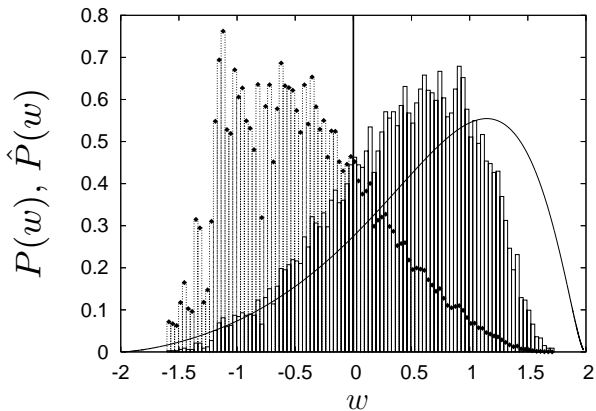}
\caption{Results for the system described by
the differential operator (\ref{FP}) with equilibrium free energy
(\ref{freeen:eq}), manipulated according to the
protocol (\ref{manh:eq}), with $J_0=J_1=0.5$, $h_0=-h_1=-1$,
and $t_\mathrm{f}=2$.  Continuous
line: probability density $P(w)$ of the work ``per spin'' $w=W/N$, with $N=2$.
The histogram of the work is obtained by 10000 simulations
of the process, see text, and is plotted with full line.
Dotted line with black diamonds: histogram of
$\hat P(w)=\exp(-\beta N w) P(w)$.
Vertical line: Thermodynamic value of the work
$w_\mathrm{rev}=\Delta F/N$. }
\label{n2}
\end{figure}

We further decrease the value of $N$ and take $N=2$, see fig.~\ref{n2}.
In this case the agreement of the histogram with the theoretical
curve is worse than
the case $N=10$, as expected.
But the small size of the system entails
a broader work distribution, and thus enables a sufficient sampling
of trajectories with $w<w_\mathrm{rev}$. In the
same figure, the histogram of the distribution
$\exp\pq{-\beta N w}P(w)$ is plotted: from this histogram
we obtain the estimate for the free energy difference
$\Delta f_\mathrm{exp}=-N^{-1}k_\mathrm{B} T \ln\left[\exp\left(-\beta
N w\right)P(N w)\right]_\mathrm{exp}$,
where $\left[\ldots\right]_\mathrm{exp}$ is the mean over
all realizations of the process. We obtain
$\Delta f_\mathrm{exp}\simeq0.015$, against a theoretical
value of $w_\mathrm{rev}=\Delta f=0$, and a most probable value of the work
$w_\mathrm{mp}\simeq 0.6$.

\section{Path thermodynamics} \label{thermo:sec}
The work distribution can be
interpreted in terms of path thermodynamics, as first suggested in
ref.~\cite{felix}. Indeed, $g(\lambda)=-\lim_{N\to\infty}\log
\Gamma(\lambda,\tf)/N$ plays the role of a path Gibbs free energy.
Thus
\begin{equation}\label{pathhelm:eq}
    \phi(w)=-\lim_{N\to\infty}\frac{1}{N}\log P(Nw,\tf),
\label{defhelm}
\end{equation}
plays the role of the corresponding Helmholtz free energy. The two
functions are related by a Legendre transformation:
\begin{eqnarray}\label{legendre:eq}
    \phi(w)&=&\inf_\lambda \left(
    g(\lambda)+\lambda w\right)\nonumber\\
    &=&g(\lambda^*(w))+\lambda^*(w) \,w,
\end{eqnarray}
where $\lambda^*(w)$ is the solution of eq.~(\ref{eqw:eq}). Thus
$\lambda$ and $w$ appear like thermodynamically conjugate
variables. Notice that if $(\lambda,w^*(\lambda))$ are a
pair of mutually conjugate variables, then $w^*(\lambda)$ is a
monotonically \emph{increasing} function of $\lambda$. Indeed
the relation between $\phi(w)$ and $\lambda$ reads
\begin{equation}
\phi'(w^*(\lambda))=\lambda.
\end{equation}
It is clear that the
most probable value of the work $w_\mathrm{mp}$ corresponds
to the value $\lambda=0$.

In ref.~\cite{felix}, $w$ is taken to play the role of the
internal energy, and thus $-\phi(w)$ that of the entropy.
Therefore $\lambda=\phi'(w)$ can be considered as an inverse
temperature. We have preferred to draw the
analogy with more familiar functions.

Indeed, one can generalize this point of view by going back to the
joint probability distribution function $\Phi(M,W,t)$. If we
define
\begin{equation}\label{bigphi:eq}
    \phi(m,w)=-\lim_{N\to\infty}\log \Phi(Nm,Nw,\tf),
\end{equation}
we obtain straightforwardly
\begin{equation}\label{biglegtr:eq}
    \phi(m,w)=\inf_{\gamma,\lambda}
    \left(\omega(\gamma,\lambda)+\gamma m+\lambda
    w\right),
\end{equation}
where $\omega(\gamma,\lambda)$ is defined in terms of
$\Omega(\lambda,\gamma,t)$, which we have defined in
eq.~(\ref{omega:def}), by
\begin{equation}\label{big:eq}
    \omega(\gamma,\lambda)=-\lim_{N\to\infty}\frac{1}{N}
    \log \Omega(\gamma,\lambda,\tf).
\end{equation}
One may notice that the $\gamma$ appearing in this equation may be
identified with $\gamma_\mathrm{f}=\gamma(\tf)$.

\section{Large fluctuations and exponential tails}
\label{expo:sec}
It was suggested in ref.~\cite{felix}, on the basis
of numerical evidence, that, for slow protocols,
the work distribution exhibits exponential tails. Here we discuss
this intriguing question.
From eq.~(\ref{pathhelm:eq}) we see that if
$P(Nw,\tf) \propto \exp (-N \lambda_0 w)$ in some interval $w_-\le w
\le w_+$, one has
\begin{equation}\label{linear:eq}
    \phi(w) = \lambda_0 w +\mbox{const.},
\end{equation}
in the same interval. A linear behavior in the Helmholtz free energy is the
signature of a first-order phase transition. In the corresponding
Gibbs free energy one has an angular point, i.e., a point
$\lambda_0$ in which
\begin{equation}\label{angular:eq}
    \lim_{\lambda\to\lambda_0^-}g'(\lambda)=w_-;
    \qquad \lim_{\lambda\to\lambda_0^+}g'(\lambda)=w_+.
\end{equation}
Thus a horizontal plateau in a plot of $\lambda^*$ vs.\ $w$ corresponds
to an exponential tail in $P(Nw,\tf)$.

We shall now follow ref.~\cite{felix}, by considering a system
of $N$ non interacting spins $\sigma_i=\pm 1$,
evolving according to the Glauber dynamics. The collective
coordinate $M$ is the total magnetization $M=\sum_i\sigma_i$, (a discrete
variable) and
the role of $\mu$ is played by the magnetic field $h$.
The system evolves according to the master equation
\begin{widetext}
\begin{eqnarray}
    \frac{\partial P}{\partial t}&=&\left\{\pd \left[\left(\frac{N+M+2}{2}\right)P(M+2,t)
    -\left(\frac{N+M}{2}\right)P(M,t)\right]\right.\nonumber\\
    &&\left.\qquad{}+\pu \left[\left(\frac{N-M+2}{2}\right)P(M-2,t)
    -\left(\frac{N-M}{2}\right)P(M,t)\right]\right\},
\label{master:eq}\end{eqnarray}ù
\end{widetext}
where the spin flip rates $p^{\uparrow,\downarrow}$ are given by
\begin{equation}
    p^\downarrow=\omega_0(h)\,\E^{-\beta h}, \qquad
    p^\uparrow=\omega_0(h)\,\E^{\beta h},
\end{equation}
in which $\omega_0(h)$ is a microscopic ``attempt
frequency'' for spin flip.
In this case, the free energy
$\FF_h(M)$ reads
\begin{eqnarray}
    \FF_h(M)&=&-h M -T S(M),
\label{Fnoni}
\end{eqnarray}
where $S(M)$ is given by eq.~(\ref{entropy:eq}) as a function of $M$.

We can make the connection with our formalism by momentarily
considering $M$ as a continuous variable, and by expressing the
shift operator
\begin{equation}\label{shift:eq}
    \mathrm T_\pm f(M)=f(M\pm 2),
\end{equation}
in the following way
\begin{equation}\label{expshift:eq}
    \mathrm T_\pm=\E^{\pm 2\frac{\partial}{\partial M}}.
\end{equation}
The master equation (\ref{master:eq}) then assumes the form
\begin{displaymath}
    \frac{\partial P}{\partial t}=\widehat \HH \,P,
\end{displaymath}
where the differential operator $\widehat \HH$ is given by
\begin{widetext}
\begin{equation}
    \widehat \HH \cdot{} =\left[\left(\E^{2\frac{\partial}{\partial  M}}
    -1\right)\left(\frac{N+M}{2}\right)
    p^\downarrow \cdot{} +\left(\E^{-2\frac{\partial}{\partial M}}-1\right)
    \left(\frac{N-M}{2}\right)p^\uparrow\cdot{} \right],
    \label{glaub}
\end{equation}
\end{widetext}
where it is understood that the derivative on $M$ acts on all instances
of $M$ it finds on its right.
Then the hamiltonian $H$, as given by eqs.~(\ref{hgm}) and
(\ref{smallH}) has the form
\begin{equation}
  H=\pq{\p{\E^{2\gamma}-1}\frac{1+m}{2} \pd
  + \p{\E^{-2\gamma}-1}\frac{1-m}{2}\pu}.
\end{equation}
Equations (\ref{eq1},\ref{eq2}) yield the equations of motion
for the classical path:
\begin{eqnarray}
    \dot m&=&  \E^{-2\gamma} \pu(1-m)-\E^{2\gamma}\pd(1+m)  ,\label{geq1}\\
    \dot \gamma&=& \frac 1 2 \pq{\p{\E^{2\gamma}-1} \pd-
    \p{\E^{-2\gamma}-1}\pu}
    -\lambda \dot h.\label{geq2}
\end{eqnarray}
A different and more complicated approach, used in ref.~\cite{felix},
leads to the same results.

We shall suppose that the
applied magnetic field $h$ is manipulated according to the simple
protocol (\ref{manh:eq})
\begin{displaymath}
h(t)=h_0+(h_1-h_0)\frac{t}{t_\mathrm{f}}, \qquad 0\le t \le t_\mathrm{f}.
\end{displaymath}
We also suppose that $\omega_0(h)=\omega_0/(\E^{\beta h}+\E^{-\beta h})$
so that the functions $\pu$, $\pd$, are explicitly given by
\begin{eqnarray}
\pu(t)&=& \omega_0\, \frac{\E^{\beta h(t)}}{\E^{\beta h(t)}
  +\E^{-\beta h(t)} },\label{pu:eq}\\
\pd(t)&=& \omega_0\, \frac{\E^{-\beta h(t)}}{\E^{\beta h(t)}
  +\E^{-\beta h(t)} }\label{pd:eq},
\end{eqnarray}
where $\omega_0$ is a constant.

\begin{figure}[ht]
\center
\psfrag{t}[ct][ct][1.]{\large $t$}
\psfrag{m}[ct][ct][1.]{\large $m_\C$}
\psfrag{0.45}[br][br][.8]{$\kappa=0.45$}
\psfrag{0.5}[br][br][.8]{$\kappa=0.5\ $}
\psfrag{0.55}[br][br][.8]{$\kappa=0.55$}
\includegraphics[width=8cm]{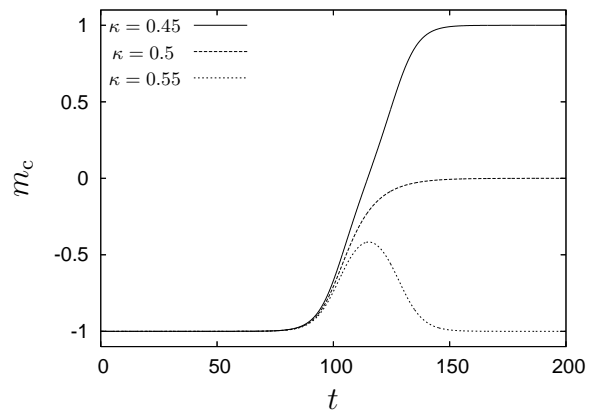}
\caption{Plot of $m_\C(t,\kappa)$ as a function of $t$,
as given by eq.~(\ref{eq:mt}),
for three values of the parameter $\kappa$.
The external field is manipulated according to eq.~(\ref{manh:eq}),
with $\tf=100$. The function is monotonic for $\kappa\le \kappa_\C$. }
\label{fig:mt}
\end{figure}

\begin{figure}[ht]
\center
\psfrag{k}[ct][ct][1.]{\large $\kappa$}
\psfrag{w}[ct][ct][1.]{\large $w$}
\includegraphics[width=8cm]{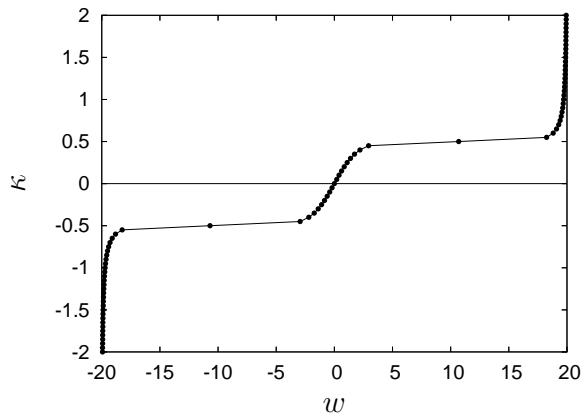}
\caption{Plot of $\kappa$ as a function of $w$ in the quasistatic
limit, as given by eq.~(\ref{wkappa}). The line is a guide to the
eye.} \label{qs_w}
\end{figure}

Let us now consider the quasi-static limit $\dot h\rightarrow 0$,
with $\lambda \dot h\rightarrow\kappa=\mathrm{const}$. It is then possible to
neglect the lhs of eqs.~(\ref{geq1},\ref{geq2}), yielding
\begin{eqnarray}
    m&=& \tanh\left(\beta h-2\gamma\right),\\
    2 \kappa&=&\pd \p{ e^{2\gamma}-1}-\pu\p{e^{-2\gamma}-1}.
\end{eqnarray}
Combining these equations, one obtains an expression for $m$
as a function of $h$:
\begin{equation}
    m_\C=\frac{ \sinh (\beta h) -2 \kappa \cosh(\beta h)}%
    {\sqrt{1+\pq{ \sinh (\beta h) -2 \kappa \cosh(\beta h)}^2}}\label{eq:mt}.
\end{equation}
Thus $m_\C$ depends on $t$ via $h$, in terms of this equation.
It also depends on the parameter $\kappa$. One can check
that $m_\C(t,\kappa)$ exhibits an extremum as a function of $t$
in the interval $\pq{0,\tf}$, if
$|\kappa|>\kappa_\C=1/2$,  otherwise
it is strictly monotonic.
In order to discuss an explicit example, we set $\beta=1$, $h_1=-h_0=10$.
In fig.~\ref{fig:mt}, the function $m_\C(t,\kappa)$,
as given by eq.~(\ref{eq:mt}) is plotted for three different
values of the parameter $\kappa$: the function clearly
exhibits a different behavior for $\kappa<\kappa_c$ and $\kappa>\kappa_c$.
Thus, as $\kappa$ becomes greater than its critical value $\kappa_c$,
we expect a singular behavior of the curve $(\kappa, w(\kappa))$, where
\begin{equation}
    w(\kappa)=-\int_0^{\tf} \D t \, \dot h(t) m_\C(t,\kappa),
\label{wkappa}
\end{equation}
and $m_\C(t,\kappa)$ is given by eq.~(\ref{eq:mt}).
Evaluating $w(\kappa)$ we obtain the curve plotted in fig.~\ref{qs_w}:
one can see that it exhibits, for $|\kappa|=\kappa_\C$, a pronounced minimum in
$\D \kappa/\D w$ rather than a horizontal plateau.

\begin{figure}[ht]
\center
\psfrag{w}[ct][ct][1.]{\large $w$}
\psfrag{k}[ct][ct][1.]{\large $\kappa$}
\includegraphics[width=8cm]{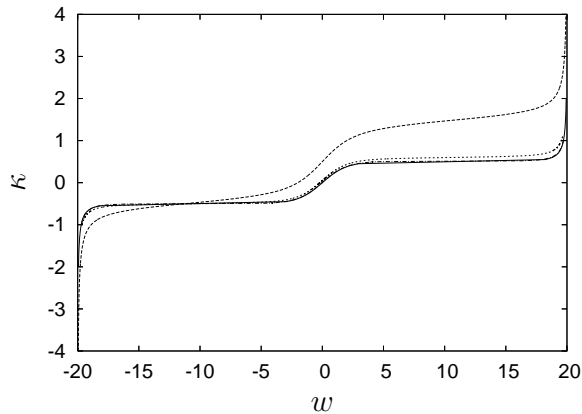}
\caption{Plot of $\kappa=\lambda \dot h$ as a function of $w$,
for different values of $\dot h=(h1-h0)/\tf$.
Thick full line: $\kappa(w)$ in the quasi-static limit,
as obtained by eq.~(\ref{wkappa}).
The other curves are obtained from the evolution equations (\ref{detpsi}).
Dashed line $\tf=20$ ($\dot h=1$), dotted line $\tf=200$ ($\dot h=0.1$),
the curve with $\tf=2000$ ($\dot h=0.01$) is practically
indistinguishable from the quasi-static limit.}
\label{q2:fig}
\end{figure}

The simplicity of the system allows us to check this prediction
by directly solving the equations (\ref{detpsi}) for the
generating function $\Psi_\sigma(\lambda,t)$, $\sigma=\pm 1$,
for the transition rates given by eqs.~(\ref{pu:eq},\ref{pd:eq}).
One thus obtains the function $\Gamma_1(\lambda,\tf)$
for the single spin, from the expression
\begin{equation}
    \Gamma_1(\lambda,\tf)=\sum_{\sigma=\pm 1}\Psi_\sigma(\lambda,\tf).
\end{equation}
Since $g(\lambda)=-\log\pq{\Gamma_1(\lambda,\tf)}$,
we can obtain the curve $(w, \lambda^*(w))$ from eq.~(\ref{eqw:eq}),
and compare it with the predicted curve for the quasi-static limit,
as given by eq.~(\ref{wkappa}).
Such a comparison is shown in fig.~\ref{q2:fig}:
as the value of $\dot h=(h_1-h_0)/\tf$ decreases the agreement
between the theory and the curve predicted by eq.~(\ref{wkappa}) improves.

\begin{figure}[ht]
\center
\psfrag{k}[ct][ct][1.]{\large $\kappa$ }
\psfrag{w}[ct][ct][1.]{\large $w$ }
\includegraphics[width=8cm]{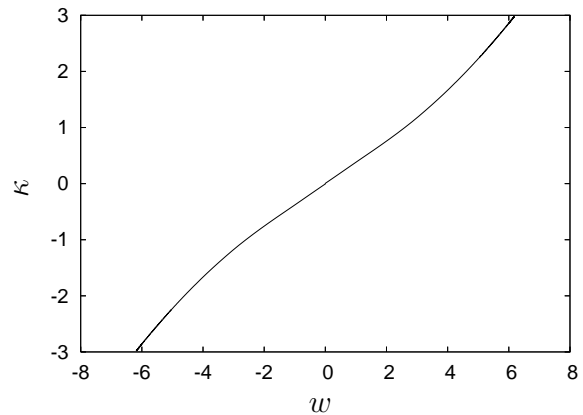}
\caption{Plot of $\kappa$ as a function of $w$ for the paramagnet
evolving according to a Langevin dynamics. Manipulation protocol
(\ref{manh:eq}), with $h_0=-h_1=-5$, $\tf=1000$.} \label{wout}
\end{figure}

We checked that this behavior depends on the details of
the dynamics by considering the same paramagnetic system,
but evolving by a Langevin rather than a Glauber dynamics,
with the method reported in section~\ref{sec3}.
As shown in fig.~\ref{wout}, the corresponding $(w,\kappa)$ curve
exhibits no plateau, and therefore
there are no exponential tails in the work distribution.
These results are confirmed by a detailed analysis of the
quasi-static limit.

Let us now turn again to the ferromagnetic mean-field system with
Langevin dynamics.

\begin{figure}[ht]
\center \psfrag{m}[ct][ct][1.]{\large $m_\C^*$} \psfrag{t}[ct][ct][1.]{\large $t$}
\psfrag{w}[ct][ct][1.]{\large $w$} \psfrag{l}[ct][ct][1.]{\large $\lambda^*$}
\includegraphics[width=8cm]{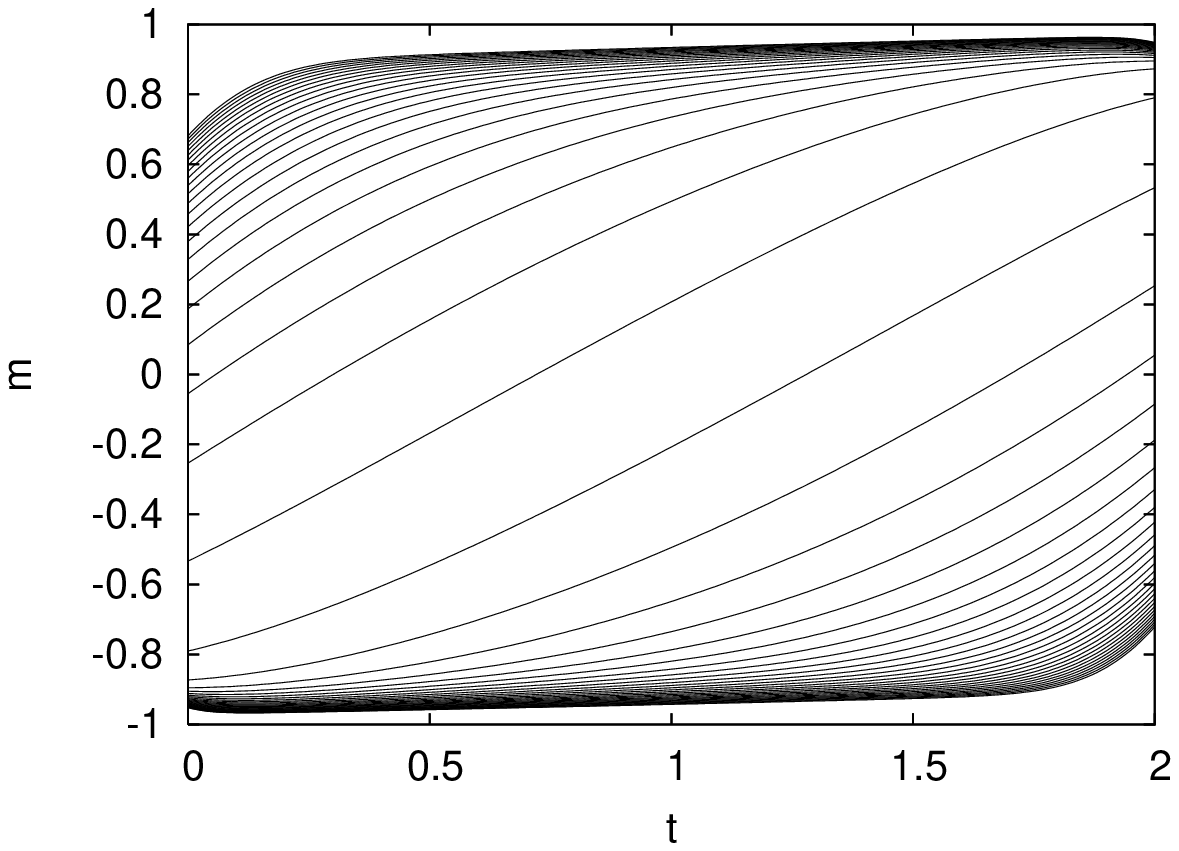}
\includegraphics[width=8cm]{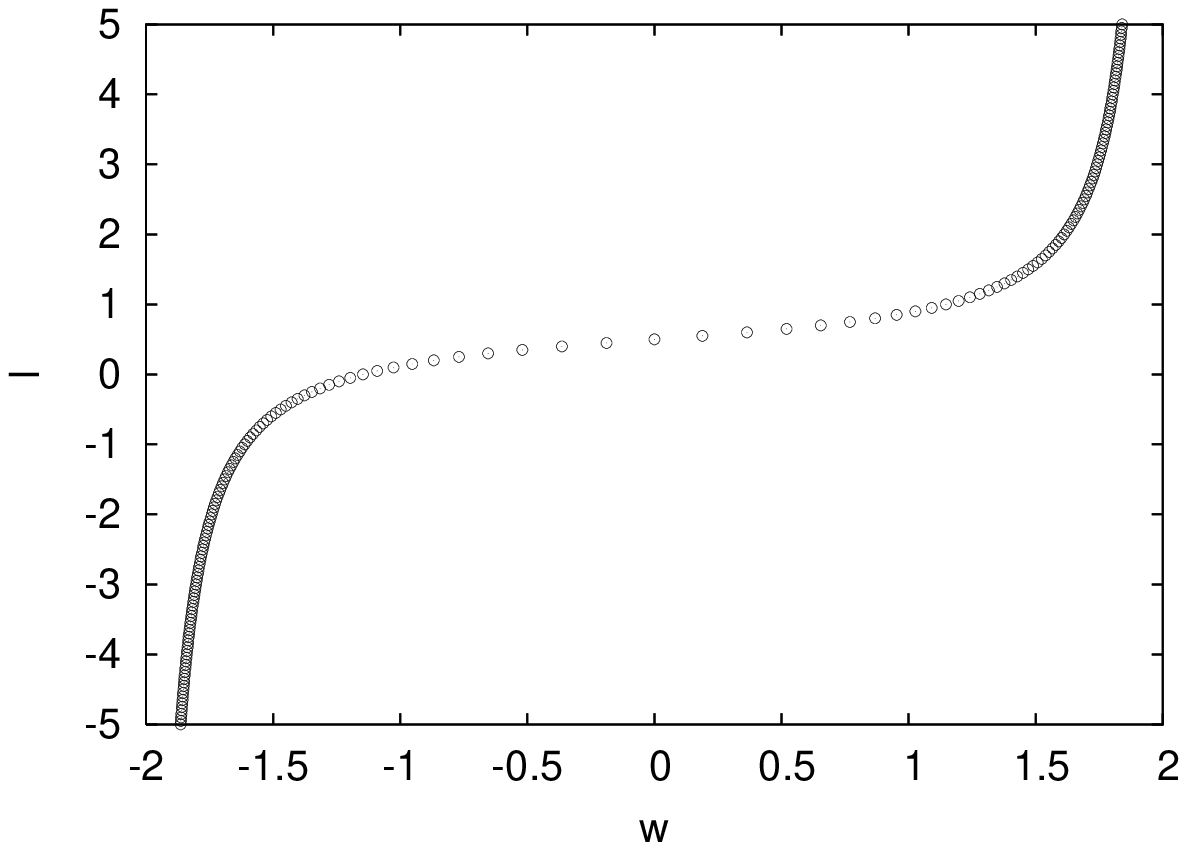}
    \caption{Top: plot of $m_\C^*$ as a function of $t$ for
    different values of $\lambda$,  with $J=0.5$, $h_0=-h_1=-1$,
and $t_\mathrm{f}=2$.
    The values of $\lambda$ vary between $\lambda=-5$
(bottom curve) and $\lambda=5$ (top curve),
    with a step  $\Delta \lambda=0.2$. Bottom:
    plot of  $\lambda^*$ as a function of $w$,
    as defined by eq.~(\ref{eqw:eq}).}
\label{mt_j0.5}
\end{figure}

In the top panel of figure \ref{mt_j0.5}, we plot $m_\C^*$ as a
function of $t$ for different values of $\lambda$, obtained by
numerical solution of eqs.~(\ref{eq1_lang},\ref{eq2_lang}), for
$J=0.5$ and $t_\mathrm{f}=2$. It can be seen that the shape of
$m_\C^*(t)$ varies continuously as $\lambda$ is varied. Accordingly
there is no horizontal plateau in the $\lambda^*$
vs.\ $w$ plot, implicitly defined by
eq.~(\ref{eqw:eq}), as shown in the bottom panel of
fig.~\ref{mt_j0.5}. The same behavior is obtained by varying $t_\mathrm{f}$
and implementing a slower or a faster protocol (data not shown).
According to the above discussion, the work distribution
exhibits no exponential tails for the case $\beta J<1$.

\begin{figure}[ht]
\center \psfrag{m}[ct][ct][1.]{\large $m_\C^*$} \psfrag{t}[ct][ct][1.]{\large $t$}
\psfrag{w}[ct][ct][1.]{\large $w$} \psfrag{l}[ct][ct][1.]{\large $\lambda^*$}
\includegraphics[width=8cm]{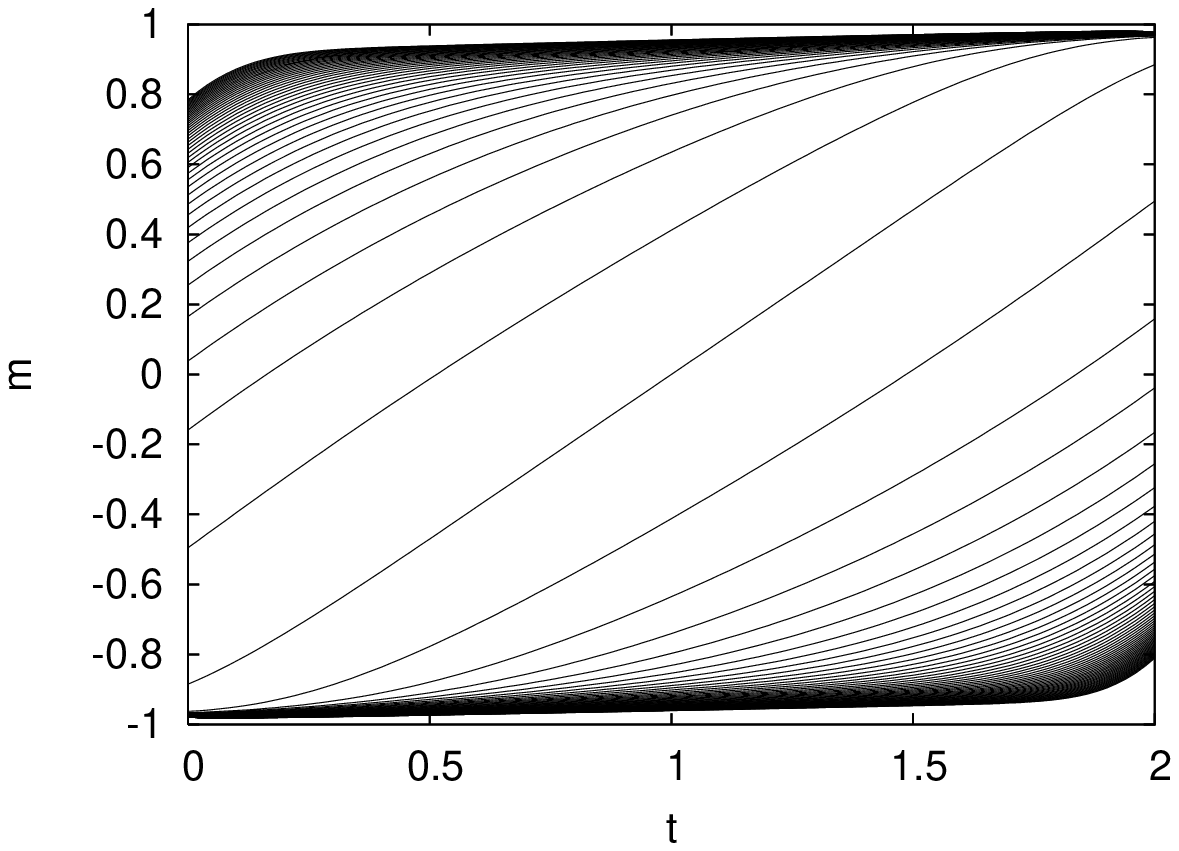}
\includegraphics[width=8cm]{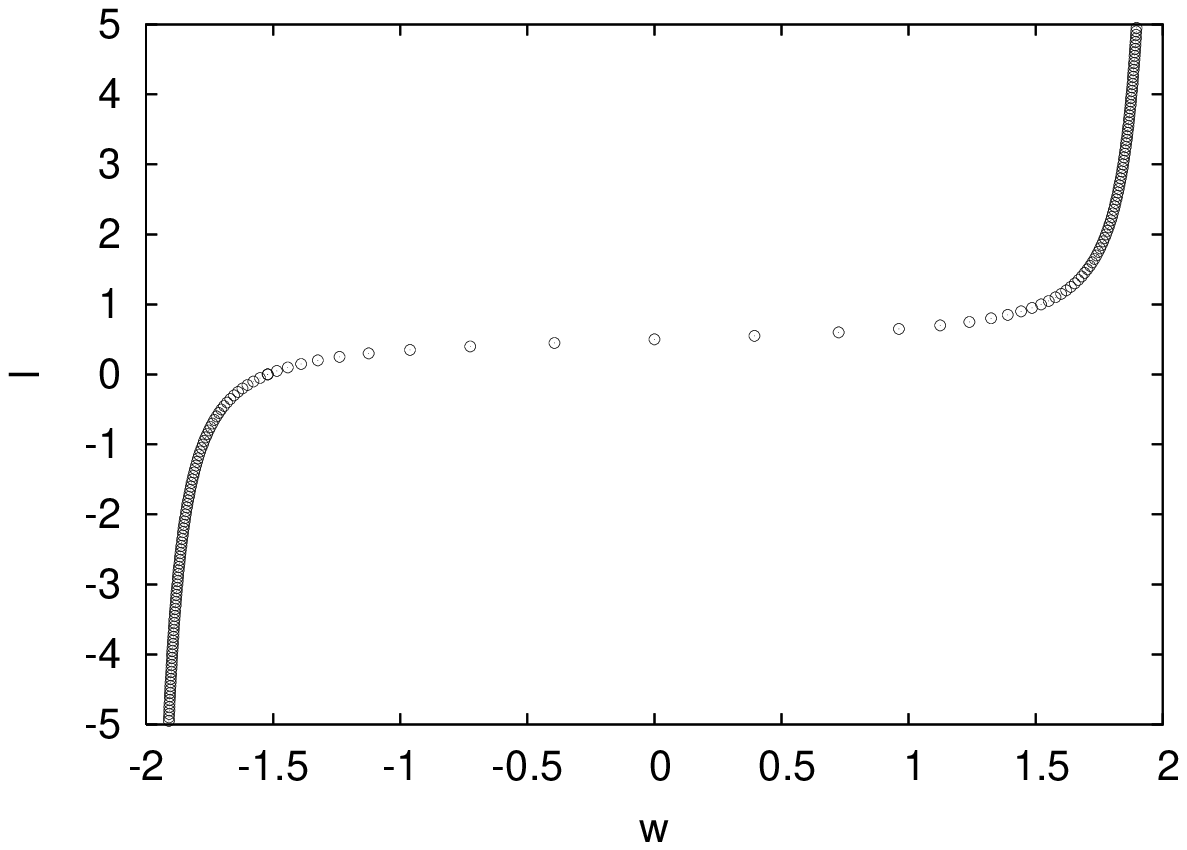}
    \caption{Top: plot of $m_\C^*$ as a function of $t$ for
    different values of $\lambda$,  with $J=1.1$, $h_0=-h_1=-1$,
    and $t_\mathrm{f}=2$.
    The values of $\lambda$ vary between $\lambda=-5$ (bottom curve)
and $\lambda=5$ (top curve),
    with a step  $\Delta \lambda=0.2$. Bottom:
plot of $\lambda^*$  as a function of  $w$,
    as defined by eq.~(\ref{eqw:eq}).}
\label{mt_j1.1}
\end{figure}

We now investigate whether a different behavior can appear when
the system is manipulated across the symmetry-breaking
transition it exhibits at $\beta J=1$. Let us look
at the behavior of the classical path $m_\C^*(t,\lambda)$
corresponding to the $\gamma_\F=0$ boundary condition, both
above ($\beta J<1$) and below ($\beta J>1$) the transition.

In the top panel of
figure~\ref{mt_j1.1}, we plot $m_\C^*$ as a function of $t$ for
different values of $\lambda$, obtained by numerical solution of
eqs.~(\ref{eq1_lang}, \ref{eq2_lang}), for $J=1.1$ and
$t_\mathrm{f}=2$: we observe no discontinuity
in $m_\C^*(t,\lambda)$ as $\lambda$ is varied,
and thus $w$ is a continuous function of $\lambda$,
see figure~\ref{mt_j1.1} bottom panel.

\begin{figure}[ht]
\center
\psfrag{m}[ct][ct][1.]{\large $m_\C^*$}
\psfrag{t}[ct][ct][1.]{\large $t$}
\psfrag{w}[ct][ct][1.]{\large $w$}
\psfrag{l}[ct][ct][1.]{\large $\lambda^*$}
\includegraphics[width=8cm]{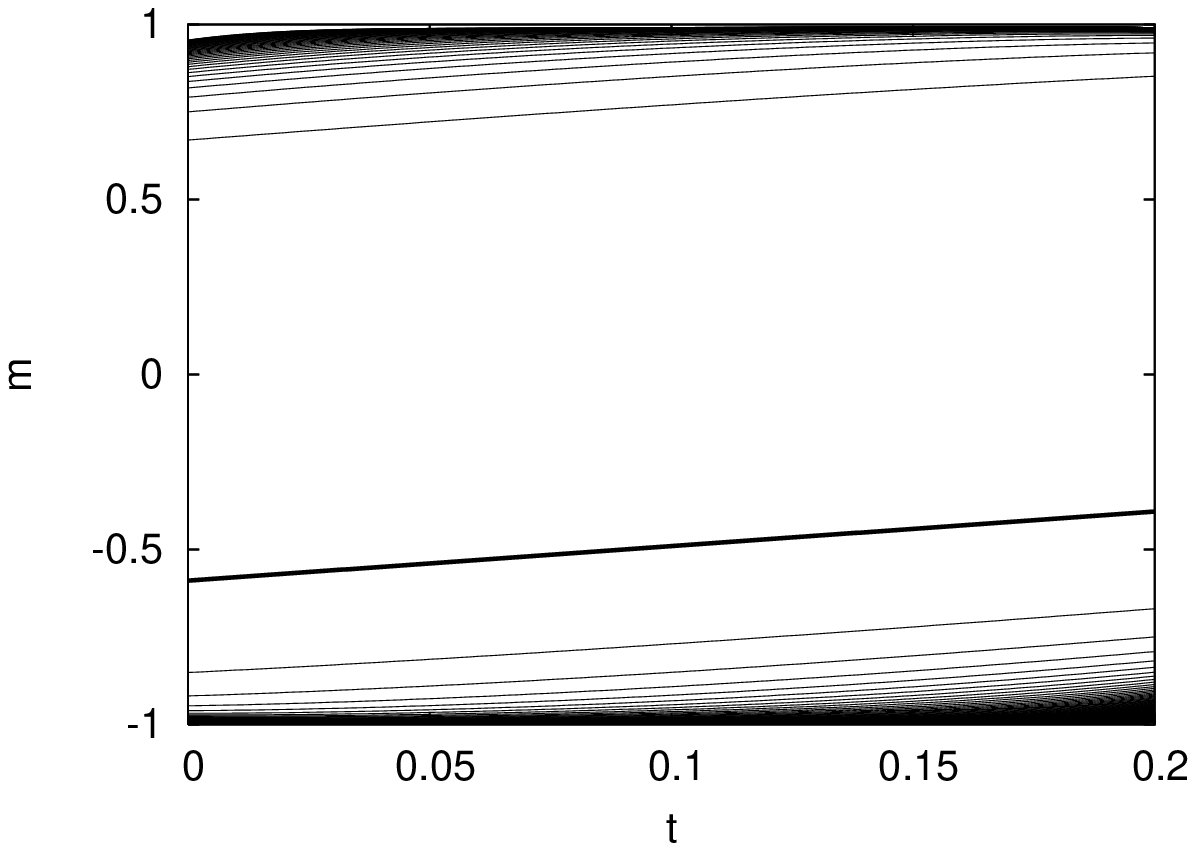}
\includegraphics[width=8cm]{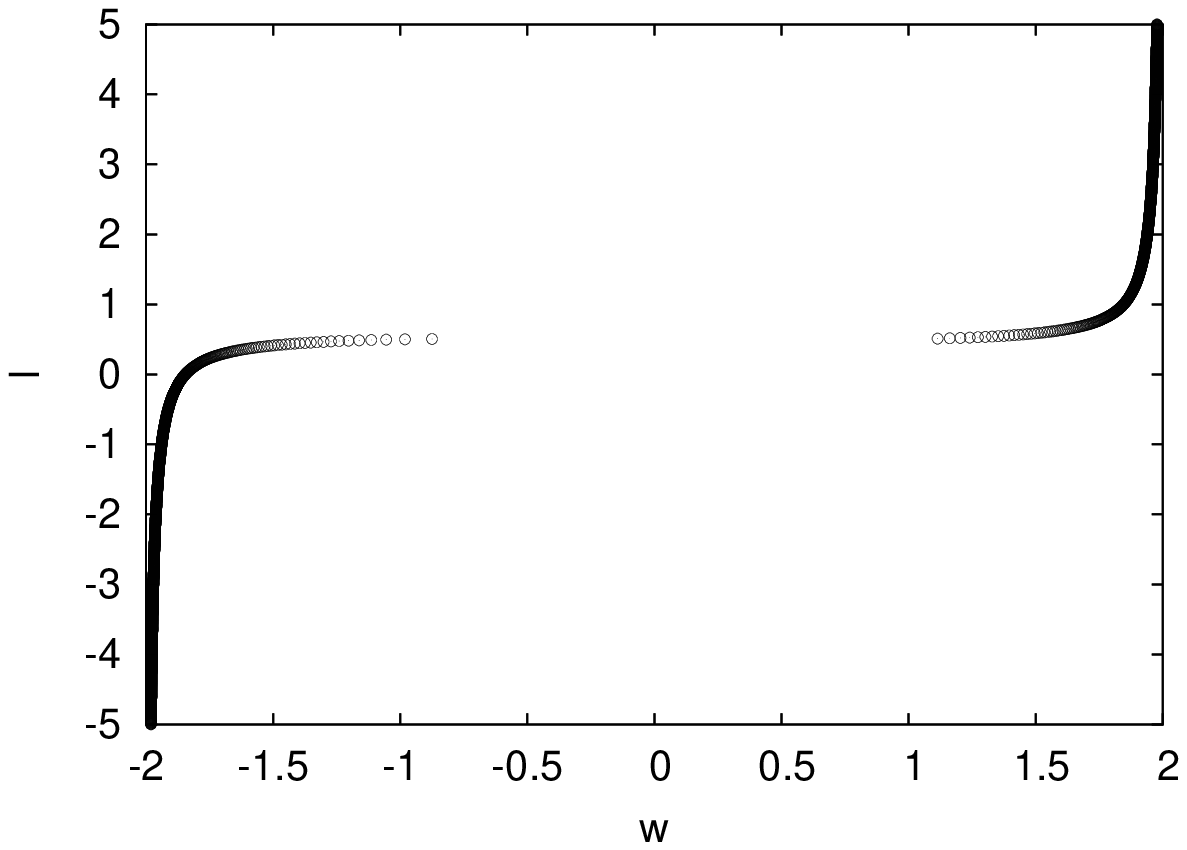}
    \caption{Top: plot of $m_\C^*$ as a function of $t$ for different
        values of $\lambda$,  with $J=1.1$, $h_0=-h_1=-1$,
        and $t_\mathrm{f}=0.2$.
        The values of $\lambda$ vary between
        $\lambda=-5$ (bottom curve) and $\lambda=5$ (top curve),
        with a step  $\Delta \lambda=0.2$.
        Thick line: $\lambda=0.5$. Bottom:  plot of $\lambda^*$ as a function
    of $w$, as defined by eq.~(\ref{eqw:eq}).}
\label{mt1_j1.1}
\end{figure}

We now consider a faster protocol, $t_\mathrm{f}=0.2$, with the same value
of $J$ and $h_0$: the results are plotted in figure \ref{mt1_j1.1}.
One can clearly see that $m_\C^*(t,\lambda)$ exhibits
a discontinuity for $\lambda=0.5$, jumping from negative to
positive values. Accordingly, $w(\lambda^*)$ exhibits a
discontinuity at $\lambda^*=0.5$, as shown in the bottom panel of
fig.~\ref{mt1_j1.1}.

\begin{figure}[ht]
\center
\psfrag{p}[ct][ct][1.]{\large $\phi$}
\psfrag{w}[ct][ct][1.]{\large $w$}
\includegraphics[width=8cm]{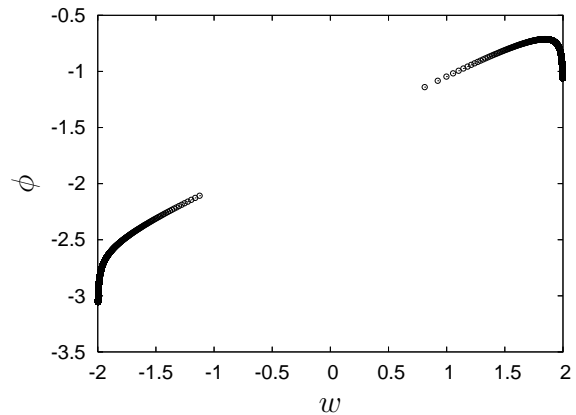}
\caption{Path Helmholtz free energy  for the system described by
the differential operator (\ref{FP}), manipulated according to the
protocol (\ref{manh:eq}), with $J=1.1$, $h_0=-h_1=-1$, and $t_\mathrm{f}=0.2$.}
\label{helm}
\end{figure}

If we now evaluate the path Helmholtz free energy $\phi(w)$ from
eq.~(\ref{legendre:eq}), we obtain the results shown in
fig.~\ref{helm}. As discussed in section~\ref{thermo:sec},
$\phi(w)$ should be obtained by a linear interpolation between
$(w_+,\phi(w_+))$ and $(w_-,\phi(w_-))$, where $w_\pm$ are the
values of $w$ either side of the discontinuity. This corresponds
to an exponential tail in the distribution of the work. In this
case, the existence of an equilibrium phase transition shows up as
a path phase transition, i.e., an exponential tail, provided that
the manipulation protocol is fast enough.
The same behavior is obtained for $h_0=-h_1=-0.1$, $t_\mathrm{f}=0.2$
(no discontinuity) and
$t_\mathrm{f}=0.02$ (discontinuity, data not shown).
This last result suggests thus that the presence of exponential tails in the
work probability distribution is due to a path ``phase separation'',
which is induced by a sufficiently fast manipulation protocol:
inspection of fig.~\ref{mt1_j1.1} indicates that the trajectories
$m_\C^*(t,\lambda)$ form two groups, as $\lambda$ is varied,
and none of the trajectories belonging to each of the two groups
crosses the line $m=0$, differently from what happens for a slower protocol,
see fig.~\ref{mt_j1.1}. We checked that the resulting distribution $\phi(w)$
satisfies the following relation, which is a consequence of Crooks' identity
\cite{Crooks} and of the symmetry $h(\tf -t)=-h(t)$ satisfied by
our protocol:
\begin{equation}
 \phi(w)-\phi(-w)=-\beta w.
\end{equation}
It would be interesting to see if such a ``path phase transition'' takes
place in more realistic models.

\section{Discussion}
In this work, we have examined the distribution of the work $W$ exerted on
a system which is manipulated out of equilibrium. We have first obtained
its expression by considering the joint distribution of
the microscopic state of the system and of the work. The expression
one obtains is in principle exact, but is amenable to numerical
solution only for very simple systems. We have then considered
a system whose quasiequilibrium state can be described by
one (or more) collective variables, to which an effective free energy
function is associated. The resulting equation for the joint
distribution of the collective variables and work is a partial differential
equation which can in principle be numerically solved. However, we
found that it is possible to explore a different direction. Indeed,
following ref.~\cite{felix}, one sees that one can
express the solution to this equation as a path integral.
In the limit of system size $N$ going to infinity, the path
integral is dominated by the classical paths, which satisfy a
``canonical'' system of ordinary differential equations, with
suitable boundary conditions. Building on this information,
it is possible to estimate the work probability distribution function
for large system size, in the form
\begin{displaymath}
  P(W)\propto \exp\left[-N\,\phi\left(w\right)\right],
\end{displaymath}
where $w=W/N$, and
$\phi(w)$ plays the role of a work free energy density, or of a function
of large deviations.  This quantity is obtained as a Legendre
transform of $g(\lambda)$ as given by eq.~(\ref{defgibbs}).
It is natural to interpret the relations between these
quantities as corresponding to those between
the Helmholtz and the Gibbs free energy densities in thermodynamics.
Within this picture,
the parameter $\lambda$ can be viewed as the intensive field
conjugated with the extensive variable $w$, which acts as an order parameter
for the single path.
Thus, horizontal plateau in the $\lambda^*$ vs.\ $w$ plot indicates
a first-order phase transition in the paths. In this case
the work distribution exhibits an exponential tail
in a given range of $w$, depending on the manipulation details.
Our results suggest that the system exhibits such path ``phase separation''
for sufficiently fast manipulation protocols, and below the mean-field
equilibrium transition temperature, whereas above it one
can find only a marked inflection point in the $\lambda^*$ vs.\ $w$ plot,
but not a horizontal plateau.

The results we obtain are interesting in their own right, since
they exhibit a number of nontrivial properties of the classical paths.
However, their usefulness for assessing the feasibility of the
use of the Jarzynski equality for the reconstruction of the
equilibrium free energy landscape can be \textit{a~priori} doubted.
Indeed, the $P(W)$ one obtains in this way is only asymptotically
valid for large $N$, and in this case, the probability of observing, in
an actual experiment, a sufficient number of large fluctuations
to evaluate the Jarzynski average (\ref{JE:eq}) with some confidence,
is extremely small. We found however that the estimated distribution
is not too far from the actual distribution for system sizes as
small as 2, at least when the manipulation protocol is not too fast
and does not cross an equilibrium phase transition line~\cite{noi2}.
In this case the JE can be
successfully applied to the work distribution obtained by simulations:
the estimate of the free energy difference
differs little from the expected value.

It is reasonable to expect, for our mean-field like systems, that
the existence of a first-order transitions could cause some problems.
Formally, in the limit $N\to\infty$ and for a manipulation protocol
with a finite speed, the system would remain close to the free-energy
minimum it finds itself in until it reaches the spinodal line. In a
finite system, if the protocol is slow enough, the system can cross
the free energy barrier and reach the real minimum in a finite time.
We found that the classical paths are able to interpolate between
the minima for slow enough protocols, whereas they tend to split
in different phases for fast ones. Thus this effect takes place
even for mean-field systems.

It is possible to extend this work to more realistic
systems, provided that the basic assumption of the existence of relevant
collective variables holds. One should also consider what information
can be gathered by exploiting other manipulation protocols.

\begin{acknowledgments}
We thank F. Ritort for his interest in our work.
This research was partially supported by MIUR-PRIN 2004.
\end{acknowledgments}

\appendix*
\section{Derivation of the Jarzynski equality for the classical paths}
\hypertarget{appendix}{}
We report here, for completeness, the derivation of the
Jarzynski equality at the level of classical paths,
obtained in ref.~\cite{noi2}.
We first show that, for $\lambda=-\beta$, the solution
of the classical equations of motion (\ref{eq1},\ref{eq2})
satisfy an equation analogous to (\ref{gamma0})
at all times, namely
\begin{equation}\label{eqQ}
    Q \equiv -\gamma-\beta\frac{\partial f_\mu}{\partial m}=0.
\end{equation}
By multiplying both sides of eq.~(\ref{equilibrium}) by
$\E^{-\gamma M}$ and integrating by parts over $M$ one
obtains
\begin{equation}
\label{saddle:eq}
\int \D M\;\HH(\gamma,M)\,\E^{-\beta\FF_\mu(M)-\gamma M}=0,
\end{equation}
where $\HH(\gamma,M)$ is given by (\ref{smallH}).
Evaluating this integral by the saddle point method
in the large $N$ limit, we obtain
\begin{equation}
\label{zeroham:eq}
H(\gamma,m^*)=0,
\end{equation}
if $\gamma$ and $m^*$ are related by
(\ref{eqQ}).
By differentiating eq.~(\ref{zeroham:eq}) with respect to $\gamma$
at fixed $\mu$ we obtain
\begin{equation}
\label{derH:eq}
\frac{\partial H}{\partial\gamma}+\left.\frac{\partial H}{\partial m}\right|_{m^*}
\left.\frac{\partial m^*}{\partial \gamma}\right)_\mu=0.
\end{equation}
Let us now take the derivative of eq.~(\ref{eqQ}) with
respect to $\gamma$ at fixed $\mu$. We obtain
\begin{equation}
\label{dergamma:eq}
\beta \left.\frac{\partial^2 f_\mu}{\partial m^2}
\frac{\partial m}{\partial \gamma}\right)_\mu=-1.
\end{equation}
By multiplying both sides of eq.~(\ref{derH:eq}) by
$\partial^2 f_\mu/\partial m^2$ and substituting
eq.~(\ref{dergamma:eq}), we obtain the following relation
\begin{equation}
\beta\frac{\partial^2 f_\mu}{\partial m^2}\frac{\partial H}{\partial \gamma}-
\frac{\partial H}{\partial m}=0,
\label{eqhf}
\end{equation}
which holds when $\gamma$ and $m$ are related by eq.~(\ref{eqQ}).
We can now evaluate the time derivative of the lhs of
 eq.~(\ref{eqQ}), when  $\gamma$ and $m$ satisfy eqs.~(\ref{eq1},\ref{eq2}). We have
\begin{eqnarray}
\dot Q &=& -\dot \gamma -\beta \frac{\partial^2 f_\mu}{\partial m^2} \dot m
-\beta \frac{\partial^2 f_\mu}{\partial m\partial \mu}\dot \mu\\
&=& -\p{\frac{\partial H}{\partial m}-\beta \frac{\partial^2 f_\mu}{\partial m\partial \mu}\dot \mu }
+\beta \frac{\partial^2 f_\mu}{\partial m^2} \frac{\partial H}{\partial \gamma}
- \beta \frac{\partial^2 f_\mu}{\partial m\partial \mu}\dot \mu.\nonumber
\end{eqnarray}
The second and the last term cancel out.
Substituting eq.~(\ref{eqhf}), we see that also
the first and the third therm cancel out.
Thus if $\gamma$ and $m$ satisfy eqs.~(\ref{eq1},\ref{eq2})
at all times, and satisfy eq.~(\ref{eqQ})
at a given time, they satisfy this last equation at any time.

Thus, for $\lambda=-\beta$, the Lagrangian, evaluated along the classical
path, is given by
\begin{eqnarray}
\LL_\mathrm{c}&=&N\left[\gamma\dot m-\beta\dot \mu \frac{\partial f_{\mu}}{\partial \mu}\right]
=N\left[-\beta \frac{\partial f_\mu}{\partial m}\dot m-\beta \frac{\partial f_\mu}{\partial \mu}\dot\mu\right]\nonumber\\
&=&-\beta N \frac{\D f_\mu}{\D t},
\end{eqnarray}
where we have exploited eq.~(\ref{eqQ}). Substituting
this expression in eq.~(\ref{integral:def}) one recovers eq.~(\ref{psimt}) and
the Jarzynski equality.

\end{document}